\documentclass[twocolumn]{aastex631}

\shorttitle{GRB 180618A}
\shortauthors{Jordana-Mitjans et al.}
\begin{document}

\title{A short gamma-ray burst from a proto-magnetar remnant}

\correspondingauthor{N\'uria Jordana-Mitjans}
\email{N.Jordana@bath.ac.uk}

\author[0000-0002-5467-8277]{N.~Jordana-Mitjans}
\affiliation{Department of Physics, University of Bath, Claverton Down, Bath, BA2 7AY, UK}

\author[0000-0003-2809-8743]{C. G.~Mundell}
\affiliation{Department of Physics, University of Bath, Claverton Down, Bath, BA2 7AY, UK}

\author[0000-0001-6869-0835]{C.~Guidorzi}
\affiliation{Department of Physics and Earth Science, University of Ferrara, via Saragat 1, I-44122, Ferrara, Italy}
\affiliation{INFN -- Sezione di Ferrara, Via Saragat 1, 44122 Ferrara, Italy}
\affiliation{INAF -- Osservatorio di Astrofisica e Scienza dello Spazio di Bologna, Via Piero Gobetti 101, 40129 Bologna, Italy}

\author[0000-0003-3434-1922]{R. J.~Smith}
\affiliation{Astrophysics Research Institute, Liverpool John Moores University, 146 Brownlow Hill, Liverpool, L3 5RF, UK}

\author[0000-0003-2558-3102]{E.~Ram\'irez-Ruiz}
\affiliation{Department of Astronomy and Astrophysics, University of California, Santa Cruz, CA 95064, USA}
\affiliation{DARK, Niels Bohr Institute, University of Copenhagen, Jagtvej 128, DK-2200, Copenhagen, Denmark}

\author[0000-0002-4670-7509]{B. D.~Metzger}
\affiliation{Columbia Astrophysics Laboratory, Columbia University, New York, New York 10027, USA}
\affiliation{Center for Computational Astrophysics, Flatiron Institute, New York, NY 10010, USA}

\author[0000-0001-7946-4200]{S.~Kobayashi}
\affiliation{Astrophysics Research Institute, Liverpool John Moores University, 146 Brownlow Hill, Liverpool, L3 5RF, UK}

\author[0000-0002-0908-914X]{A.~Gomboc}
\affiliation{Center for Astrophysics and Cosmology, University of Nova Gorica, Vipavska 13, 5000 Nova Gorica, Slovenia}

\author[0000-0001-8397-5759]{I. A.~Steele}
\affiliation{Astrophysics Research Institute, Liverpool John Moores University, 146 Brownlow Hill, Liverpool, L3 5RF, UK}

\author[0000-0002-4022-1874]{M. Shrestha}
\affiliation{Astrophysics Research Institute, Liverpool John Moores University, 146 Brownlow Hill, Liverpool, L3 5RF, UK}

\author[0000-0002-5817-4009]{M.~Marongiu}
\affiliation{INAF -- Osservatorio Astronomico di Cagliari - via della Scienza 5 - I-09047 Selargius, Italy}

\author[0000-0002-8860-6538]{A.~Rossi}
\affiliation{INAF -- Osservatorio di Astrofisica e Scienza dello Spazio, via Piero Gobetti 93/3, 40129 Bologna, Italy}

\author[0000-0003-2283-2185]{B.~Rothberg}
\affiliation{LBT Observatory, University of Arizona, 933 N.Cherry Ave,Tucson AZ 85721, USA}
\affiliation{George Mason University, Department of Physics \& Astronomy, MS 3F3, 4400 University Drive, Fairfax, VA 22030, USA}

%%%%%%%%%%%%%%%%%%%%%%%%%%%%%%%%%%%%%%%%%%%%%%%%%%%%%%%%%%%%%%%%%%%%%%%%%%%%%%%%

\begin{abstract}
The contemporaneous detection of gravitational waves and gamma rays from the GW170817/GRB 170817A, followed by kilonova emission a day after, confirmed compact binary neutron-star mergers as progenitors of short-duration gamma-ray bursts (GRBs), and cosmic sources of heavy r-process nuclei. However, the nature (and lifespan) of the merger remnant and the energy reservoir powering these bright gamma-ray flashes remains debated, while the first minutes after the merger are unexplored at optical wavelengths. Here, we report the earliest discovery of bright thermal optical emission associated with the short GRB 180618A with extended gamma-ray emission ---with ultraviolet and optical multicolour observations starting as soon as $1.4\,$minutes post-burst. The spectrum is consistent with a fast-fading afterglow and emerging thermal optical emission at $15\,$minutes post-burst, which fades abruptly and chromatically (flux density $F_{\nu} \propto t^{-\alpha}$, $\alpha=4.6 \pm 0.3$) just $35 \,$minutes after the GRB. Our observations from gamma rays to optical wavelengths are consistent with a hot nebula expanding at relativistic speeds, powered by the plasma winds from a newborn, rapidly-spinning and highly magnetized neutron star (i.e. a millisecond magnetar), whose rotational energy is released at a rate $L_{\rm th} \propto t^{-(2.22\pm 0.14)}$ to reheat the unbound merger-remnant material. These results suggest such neutron stars can survive the collapse to a black hole on timescales much larger than a few hundred milliseconds after the merger, and power the GRB itself through accretion. Bright thermal optical counterparts to binary merger gravitational wave sources may be common in future wide-field fast-cadence sky surveys. 
\end{abstract}


%%%%%%%%%%%%%%%%%%%%%%%%%%%%%%%%%%%%%%%%%%%%%%%%%%%%%%%%%%%%%%%%%%%%%%%%%%%%%%%%

\section{Introduction} \label{sec:intro}

Gamma-ray bursts (GRBs) are bright extragalactic flashes of gamma-ray radiation and briefly the most energetic explosions in the Universe \citep{2009ARA&A..47..567G}. Their catastrophic origin ---the merger of compact star binaries for short-duration GRBs \citep{1986ApJ...308L..43P,1999ApJ...520..650F,2013Natur.500..547T,2007NJPh....9...17L,2017ApJ...848L..13A} or the collapse of massive stars for long GRBs \citep{1993ApJ...405..273W,1999ApJ...524..262M,1999Natur.401..453B}--- drives the formation of a newborn compact remnant (black hole or magnetar) that powers two highly relativistic jets. In the framework of the standard fireball model and after the initial prompt gamma-ray emission (e.g., \citealt{1997ApJ...476..232M,1999PhR...314..575P}), the relativistic ejecta are decelerated by the circumburst medium by a pair of external shocks: a short-lived reverse shock and a forward shock (e.g., \citealt{1992MNRAS.258P..41R,1999ApJ...520..641S, 2000ApJ...545..807K}). This lagging emission called the afterglow radiates via synchrotron emission and can be detected seconds to years after the burst at wavelengths across the electromagnetic spectrum (e.g., \citealt{1997Natur.387..783C,1997Natur.386..686V,2009ARA&A..47..567G}).

Short GRBs represent the $9\%$ of the total detected by the Swift Burst Alert Telescope (BAT; \citealt{2016ApJ...829....7L}), resulting in a significantly lower frequency of real-time multiwavelength studies when compared to long GRBs. Additionally, the optical counterparts of short GRBs are typically a few hundred times fainter than those of massive star collapse origin \citep{2011ApJ...734...96K}. This challenges the early follow-up and the study of short GRBs with small and medium-sized telescopes, and limits the available data to several hours post-burst, in the kilonovae time domain \citep{2013Natur.500..547T,2017ApJ...848L..12A,2018NatCo...9.4089T}.

Successful broadband follow-up of short GRBs began with the discovery of the X-ray and optical afterglow of the GRB 050709 \citep{2005Natur.437..855V,2005Natur.437..845F,2005Natur.437..859H}, and first radio afterglow of the GRB 050724 \citep{2005Natur.438..988B}. After these events, there have been numerous detections of short GRB afterglows (e.g., \citealt{2015ApJ...815..102F}), including the first detection of a kilonova \citep{2013Natur.500..547T,2013ApJ...774L..23B} and the joint discovery of the GW170817/GRB 170817A/kilonova, which confirmed that binary neutron stars are progenitors of at least some short GRBs \citep{2017ApJ...848L..12A,2017ApJ...848L..13A,2017ApJ...848L..27T,2017ApJ...848L..14G}. Still, the remnant of neutron star binary mergers remains largely debated \citep{2014ApJ...788L...8M,2021ApJ...908..152M, 2017PhRvD..96h4063R,2018ApJ...857...95M, 2019LRR....23....1M,2019ApJ...880L..15M, 2020ApJ...901L..37M,2020ApJ...888...97B}. More recently, giant flares from extragalactic magnetars have been associated as sources of low-luminosity short-duration GRBs \citep{2021Natur.589..207R,2021Natur.589..211S,2021NatAs.tmp...11F}.

Here, we present the early-time multiwavelength observations and polarization constraint of the GRB 180618A, a short GRB with extended emission (e.g., \citealt{2006ApJ...643..266N}).  So far, the only polarization measurement of a short GRB optical counterpart has been the $P = 0.50 \% \pm 0.07\%$ detection in the GW170817 kilonova at $\approx 1.5\,$days after the merger \citep{2017NatAs...1..791C} ---consistent with polarization from the Galactic dust. Short GRBs with extended soft gamma-ray emission are rarely studied at lower frequencies \citep{2009ApJ...696.1871P,2017A&A...607A..84K}, as they are a small fraction of the total detected by the BAT ($\approx 1 \%$; \citealt{2016ApJ...829....7L}). Such elusive objects are merger candidates and display the typical short-hard prompt gamma-ray emission followed by variable soft gamma-ray emission spanning 10$-$100$\,$s \citep{2006ApJ...643..266N,2009ApJ...696.1871P,2014ApJ...789..145H,2015MNRAS.452..824K}. Candidate mechanisms powering such extended gamma-ray emission after the merger include late-time activity from the central engine \citep{2008MNRAS.385.1455M, 2012MNRAS.419.1537B}, interaction with a pulsar-wind cavity \citep{2019ApJ...883L...6R}, prolonged accretion from the gravitationally-bound material ejected pre-merger \citep{2007MNRAS.376L..48R,2007NJPh....9...17L,2009ApJ...699L..93L}, or a two-component outflow viewed slightly off-axis \citep{2011MNRAS.417.2161B}.

This work is structured as follows. In Section \ref{sec:methods}, we present the GRB 180618A optical observations and data reduction of the UltraViolet and Optical Telescope, and the 2-m Liverpool Telescope ---including the RINGO3 multiwavelength polarimeter/imager and the IO:O camera. As well, we detail the 8.4-m Large Binocular Telescope observations; that is, deep-field imaging with the Large Binocular Cameras and spectroscopy of the host galaxy candidates with the Multi-Object Double Spectrographs. In Section \ref{sec:results}, the properties of the optical and gamma-ray emission are presented. In Section \ref{sec:inter}, the physical origin of the peculiar multiwavelength emission of the GRB 180618A is discussed ---in particular, the optical emission. In Section \ref{sec:discuss}, the implications of the GRB 180618A results are discussed in the wide context of neutron star mergers. In Section \ref{sec:conclusion}, we summarize our findings. 

We assume flat $\Lambda$CDM cosmology $\Omega_m = 0.32$, $\Omega_{\Lambda} = 0.68$ and $h=0.67$, as reported by \cite{2020A&A...641A...6P}. We adopt the convention $F_{\nu} \propto t ^{-\alpha} \nu ^{-\beta} $, where $\alpha$ is the temporal index and $\beta$ is the spectral index. Note that the spectral index is related to the photon index like $\beta=\beta_{\rm PI}-1$. Unless stated otherwise, all uncertainties reported in this paper are given at $1\sigma$ confidence level.

%%%%%%%%%%%%%%%%%%%%%%%%%%%%%%%%%%%%%%%%%%%%%%%%%%%%%%%%%%%%%%%%%%%%%%%%%%%%%%%%

\section{Observations and Data Reduction} \label{sec:methods}

\begin{figure}[]
\centering
\includegraphics[width=\columnwidth]{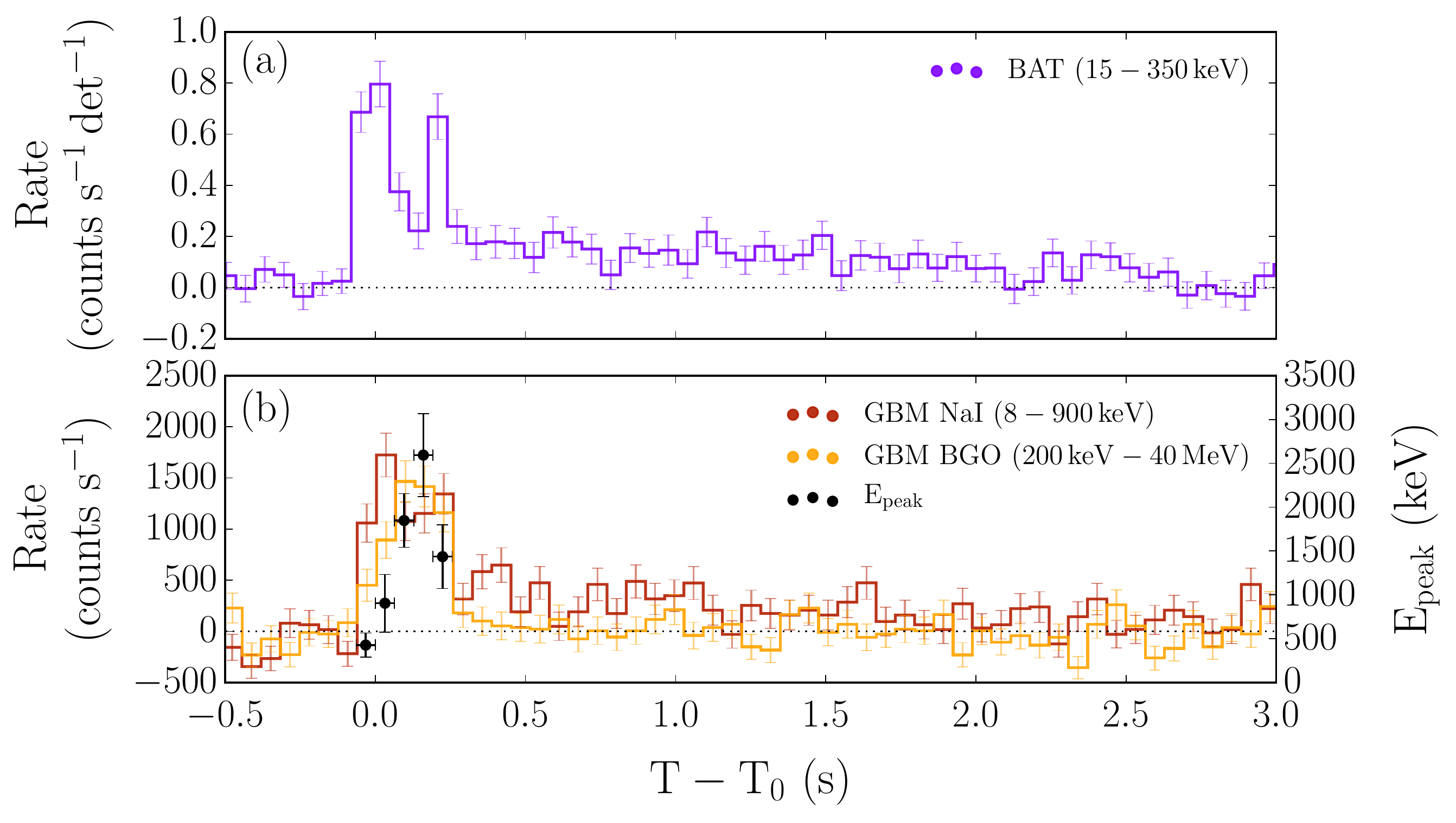}
\caption{The GRB 180618A light curves at $64\,$ms resolution as detected by the BAT and the GBM instruments. (a) Count rate per enabled detector of the 15$-$350 keV energy band of the BAT. (b) Count rates of the 8$-$900 keV energy band of the GBM sodium iodide (NaI) detector, and the 200 keV$-$40 MeV of the GBM bismuth germanate (BGO) detector. In a different y-axis, we present the evolution of the peak energy ($E_{\rm peak}$); the values were derived from a cutoff power-law model fit to the 8 keV$-$40 MeV $\nu F_{\nu}$ spectrum. In the x-axis, the T$_0$ corresponds to the BAT trigger time.}
\label{fig:LC_180618A_gamma}
\end{figure}

On 18 June 2018 at T$_0$=00:43:13 Universal Time (UT), the BAT from the Neil Gehrels Swift Observatory triggered an alert for the GRB 180618A \citep{2009ApJ...702..791M,2018GCN.22790....1L}. The GRB 180618A was detected by the BAT \citep{2005SSRv..120..143B,2018GCN.22796....1S}, the Gamma-ray Burst Monitor (GBM; \citealt{2009ApJ...702..791M,2018GCN.22794....1H}), and the Konus instrument from the Wind satellite \citep{1995SSRv...71..265A,2018GCN.22822....1S} as a short-duration and spectrally-hard bright GRB with a long-duration weak emission tail at low gamma-ray energies (see Figure~\ref{fig:LC_180618A_gamma}). Further GRB 180618A detections include the Astrosat Cadmium Zinc Telluride Imager (CZTI; \citealt{2014SPIE.9144E..1SS,2018GCN.22842....1S}), and the Insight Hard X-ray Modulation Telescope (HXMT; \citealt{2020SCPMA..6349503L,2022ApJS..259...46S}).

\subsection{Ultraviolet and Optical Light Curves} \label{sec:optreduc}

\begin{figure*}
\plotone{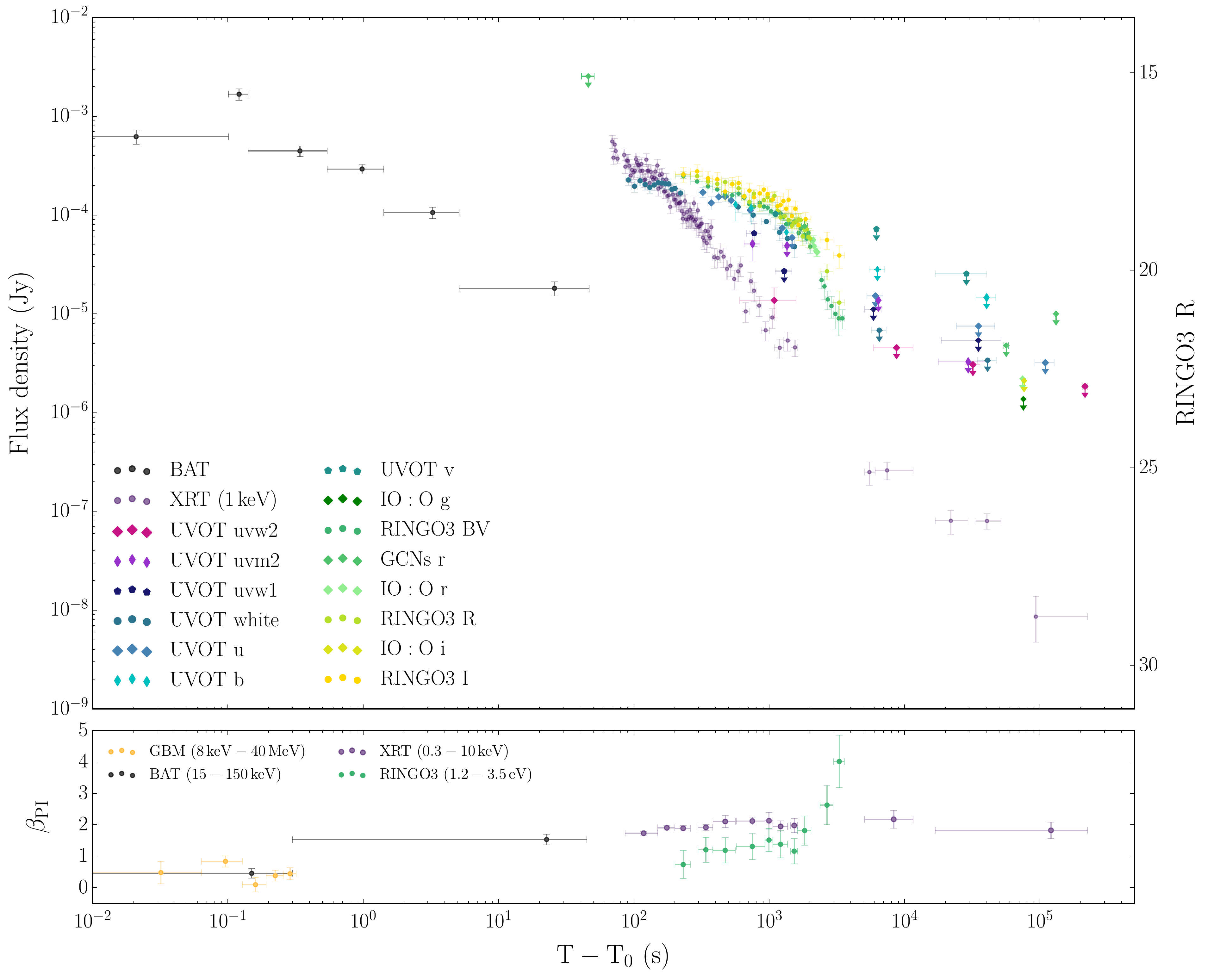}
\caption{The GRB 180618A light curves at gamma-ray, X-ray, ultraviolet and optical bands. The data corresponds to the Swift BAT, Swift XRT, Swift UVOT {\it white,v,b,u,uvw1,uvm2,uvw2} bands, LT RINGO3 {\it BV,R,I} bands and LT IO:O {\it g,r,i} bands. The Swift BAT and XRT observations were obtained from the web interface provided by the Leicester University \citep{2009MNRAS.397.1177E}; the BAT data were binned to a signal-to-noise of 7 and the absorbed 0.3$-$10 keV XRT light curve was converted to observed flux density at 1 keV. For completeness, we include the optical observations and upper limits reported in the Gamma-ray Coordination Network (GCN) from the MASTER II \citep{2018GCN.22797....1T}, Tien Shan Astronomical Observatory \citep{2018GCN.22809....1M} and Xinglong-2.16m \citep{2018GCN.22804....1Z}. Note that the GCN observations do not include filter corrections. In the x-axis, the T$_0$ corresponds to the BAT trigger time. In the y-axis, the flux density is converted to the RINGO3 {\it R} magnitude. Detections have $1\sigma$ error bars, and non-detections are presented as $3\sigma$ upper limits ---note that the MASTER data are $5\sigma$ upper limits. In the bottom panel, we present the temporal evolution of the photon index in the GBM 8 keV$-$40 MeV gamma-ray band, the BAT 15$-$150 keV gamma-ray band, the XRT 0.3$-$10 keV X-ray band, and the RINGO3 1.2$-$3.5 eV optical band.}
    \label{fig:LC_GRB180618A}
\end{figure*}

At $86\,$s after the detection of the GRB 180618A by the BAT, the UltraViolet and Optical Telescope  (UVOT; \citealt{2005SSRv..120...95R}) from the Neil Gehrels Swift Observatory started optical observations in an unfiltered band ({\it white}). Subsequently, the UVOT continued observations with the {\it uvw1,uvm2,uvw2} ultraviolet and the {\it v,b,u} optical filters (see Figure~\ref{fig:LC_GRB180618A}). At $202.5 \,$s after the BAT alert, the 2-m fully robotic Liverpool Telescope (LT; \citealt{2004AN....325..519S}) ---with site at Roque de Los Muchachos Observatory (ORM, Spain)--- automatically started follow-up observations \citep{2006PASP..118..288G} with the RINGO3 three-band polarimeter and imager \citep{2012SPIE.8446E..2JA}. The LT observations consisted of three consecutive observing sequences of $10\,$minutes each with the RINGO3 in three simultaneous bands ({\it BV,R,I}), followed by six single $10 \,$s exposures with the $r$ filter of the IO:O optical widefield camera\footnote{\url{https://telescope.livjm.ac.uk/TelInst/Inst/IOO/}}, and two extra observing sets of $10 \,$minutes with the RINGO3. Three $300$-s exposures with the IO:O {\it g,r,i} filters were also scheduled via the LT {\it phase2UI}\footnote{\url{https://telescope.livjm.ac.uk/PropInst/Phase2/}}, and autonomously executed by the LT at $7.4 \times 10^4 \,$s post-burst.

\begin{deluxetable}{ccccccccc}[]
\tablecaption{The GRB 180618A ultraviolet and optical photometry corresponding to the Swift UVOT {\it white,v,b,u,uvw1,uvm2}, {\it uvw2} bands, LT RINGO3 {\it BV,R,I} bands and LT IO:O {\it g,r,i} bands.
 \label{tab:phot}}
\tablecolumns{8}
\tablewidth{0pt}
\tablehead{
\colhead{Band} & \colhead{t$_{\rm mid}-$T$_{0}$} & \colhead{t$_{\rm exp}/2$}  & \colhead{mag} & \colhead{mag$_{\rm \, err}$}  & \colhead{F$_{\nu}$} & \colhead{F$_{\nu \, \rm{err}}$} \\
\colhead{} & \colhead{(s)} & \colhead{(s)} & \colhead{} & \colhead{} & \colhead{(Jy)} & \colhead{(Jy)} }
\startdata
{\it white} &   91 &    5 & 17.32 &  0.13 & 2.27e-04 & 2.7e-05\\
{\it white} &  101 &    5 & 17.48 &  0.14 & 1.96e-04 & 2.6e-05\\
{\it white} &  111 &    5 & 17.34 &  0.13 & 2.23e-04 & 2.6e-05\\
{\it white} &  121 &    5 & 17.44 &  0.13 & 2.04e-04 & 2.5e-05\\
{\it white} &  131 &    5 & 17.51 &  0.14 & 1.91e-04 & 2.5e-05\\
{\it white} &  141 &    5 & 17.45 &  0.13 & 2.02e-04 & 2.5e-05\\
{\it white} &  151 &    5 & 17.39 &  0.13 & 2.13e-04 & 2.5e-05\\
{\it white} &  161 &    5 & 17.40 &  0.13 & 2.11e-04 & 2.5e-05\\
{\it white} &  ... & ... & ... & ... & ... & ...
\enddata
\tablecomments{The t$_{\rm mid}$ corresponds to the mean observing time, the T$_{0}$ is the BAT trigger time, and the t$_{\rm exp}$ is the length of the observing time window. Note that the photometry is not corrected by either Galactic, i.e. with reddening E(${B-V}$)$_{\rm MW} = 0.065 \pm 0.003$ \citep{1998ApJ...500..525S}, or host galaxy extinction. Table \ref{tab:phot} is published in its entirety in machine-readable format. A portion is shown here for guidance regarding its form and content.}
\end{deluxetable}

\begin{figure}[ht!]
\centering
\includegraphics[width=\columnwidth]{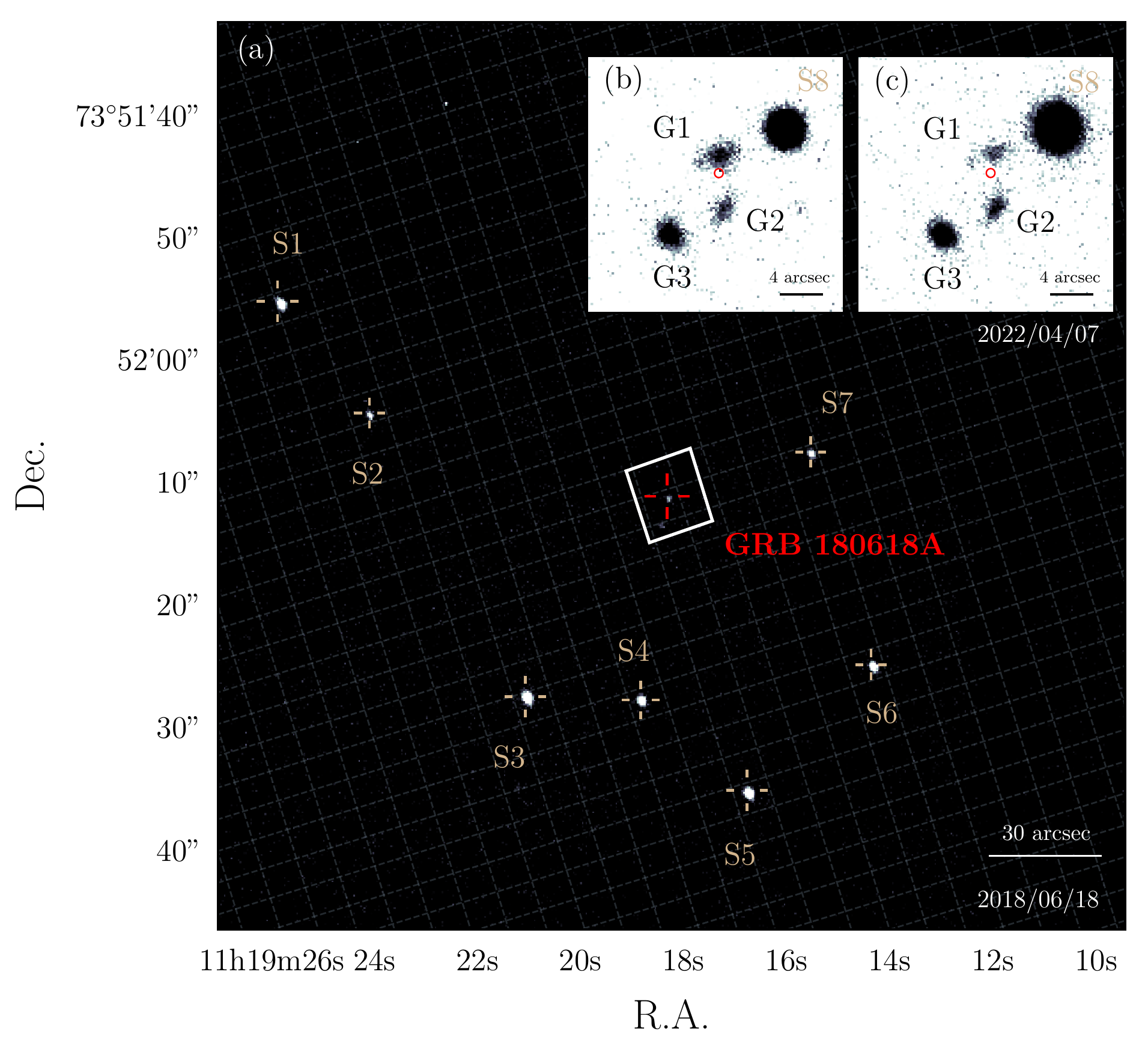}
\caption{Observing fields of the GRB 180618A sky region. The GRB 180618A location is labelled in red, the stars (S) in light brown, and the galaxies (G) in black. (a) The 2-m LT $I$-band RINGO3 image of the GRB 180618A optical transient. (b) The $r$-band LBC image from the 8.4-m LBT. (c) The LBT $z$-band image. The field of view of the magnified LBT images corresponds to the white rectangle of panel a. The LBT images reveal three galaxies at a similar redshift near the UVOT sub-arcsec localization of the GRB 180618A (shown in red at $90\%$ confidence level). The G1 is the host galaxy of the short GRB 180618A, with spectroscopic redshift $z=0.554 \pm 0.001$.}
\label{fig:fov}
\end{figure}

\begin{figure*}[ht!]
\centering
\includegraphics[width=59mm]{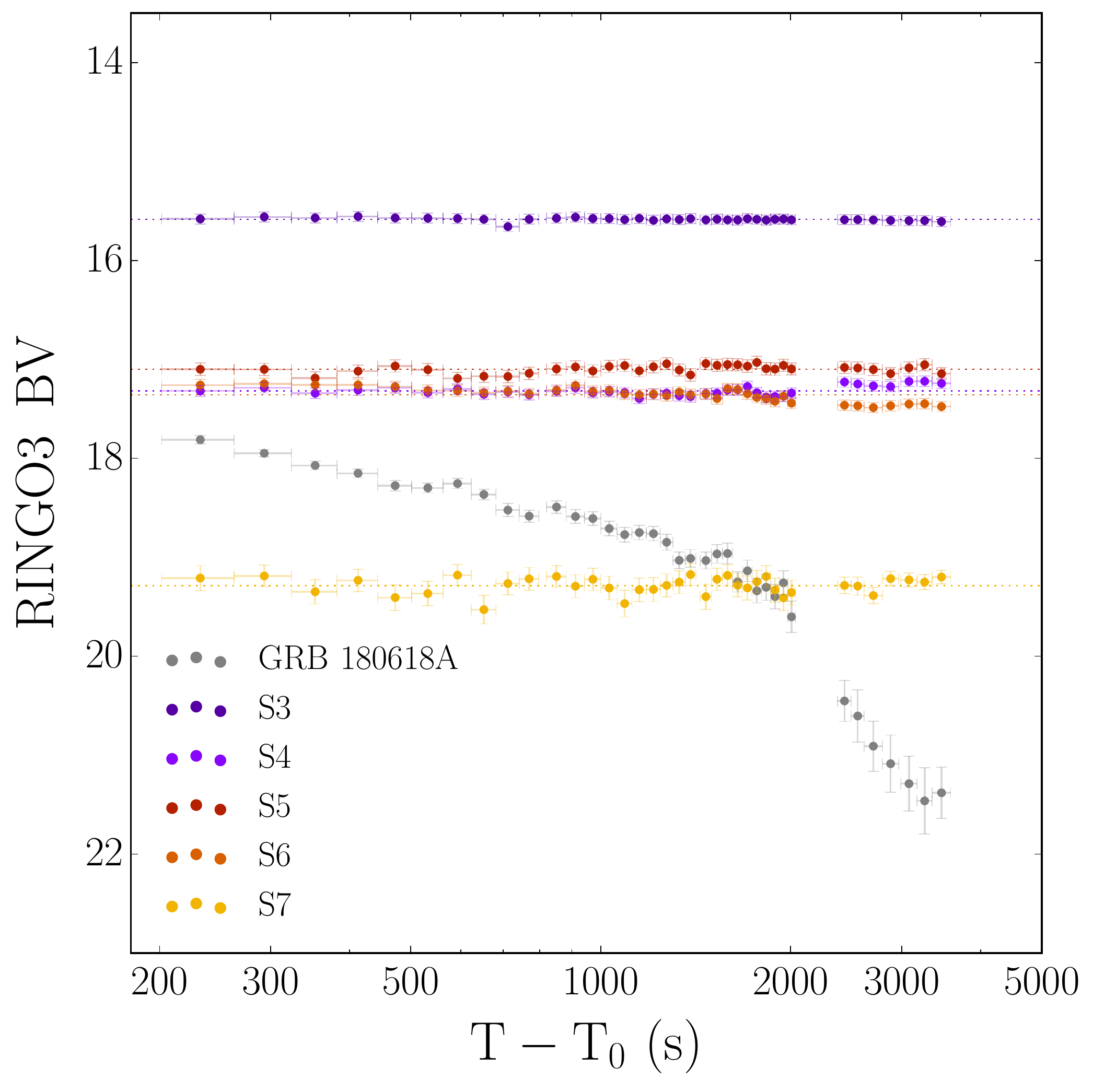}
\includegraphics[width=59mm]{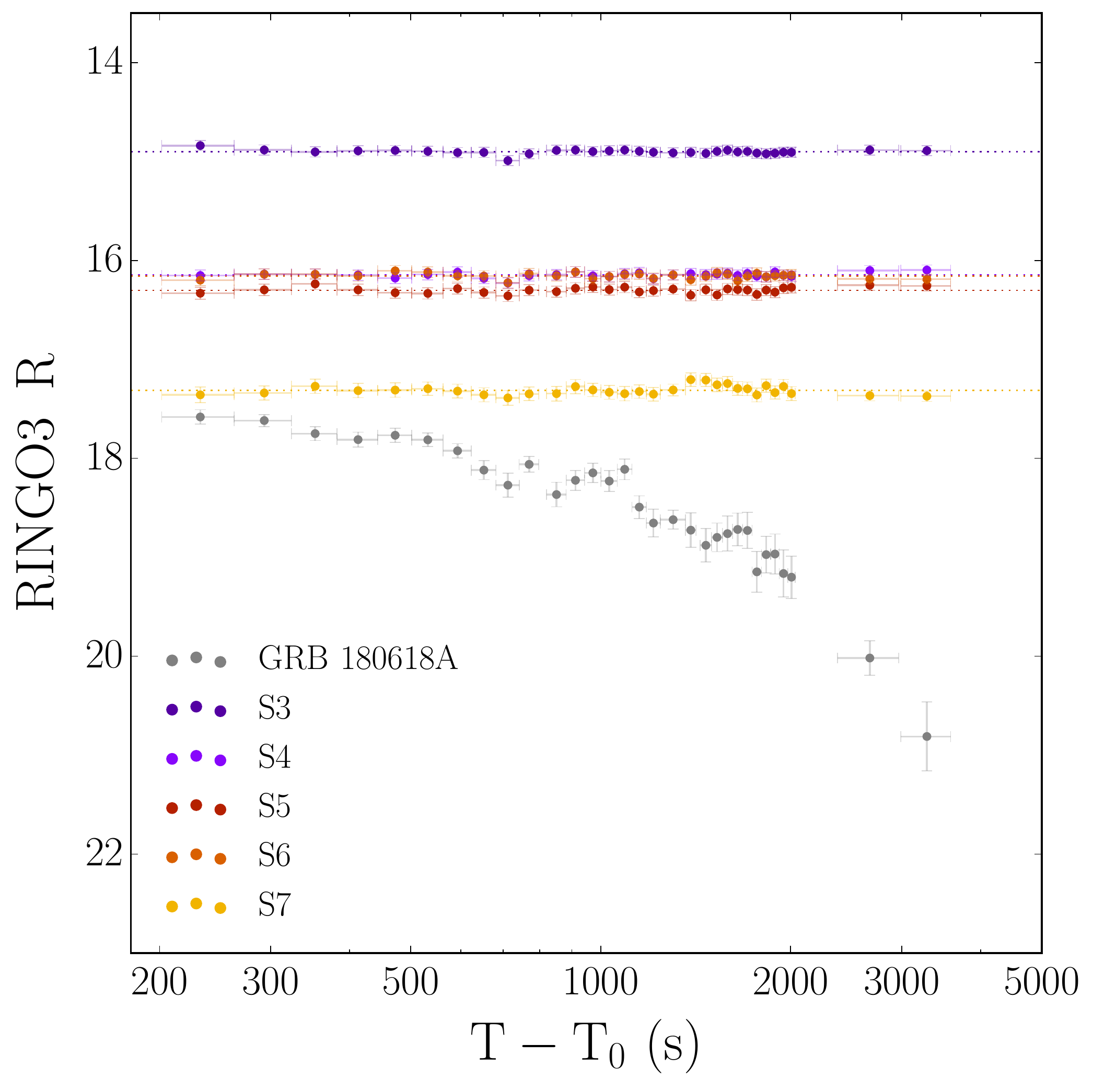}
\includegraphics[width=59mm]{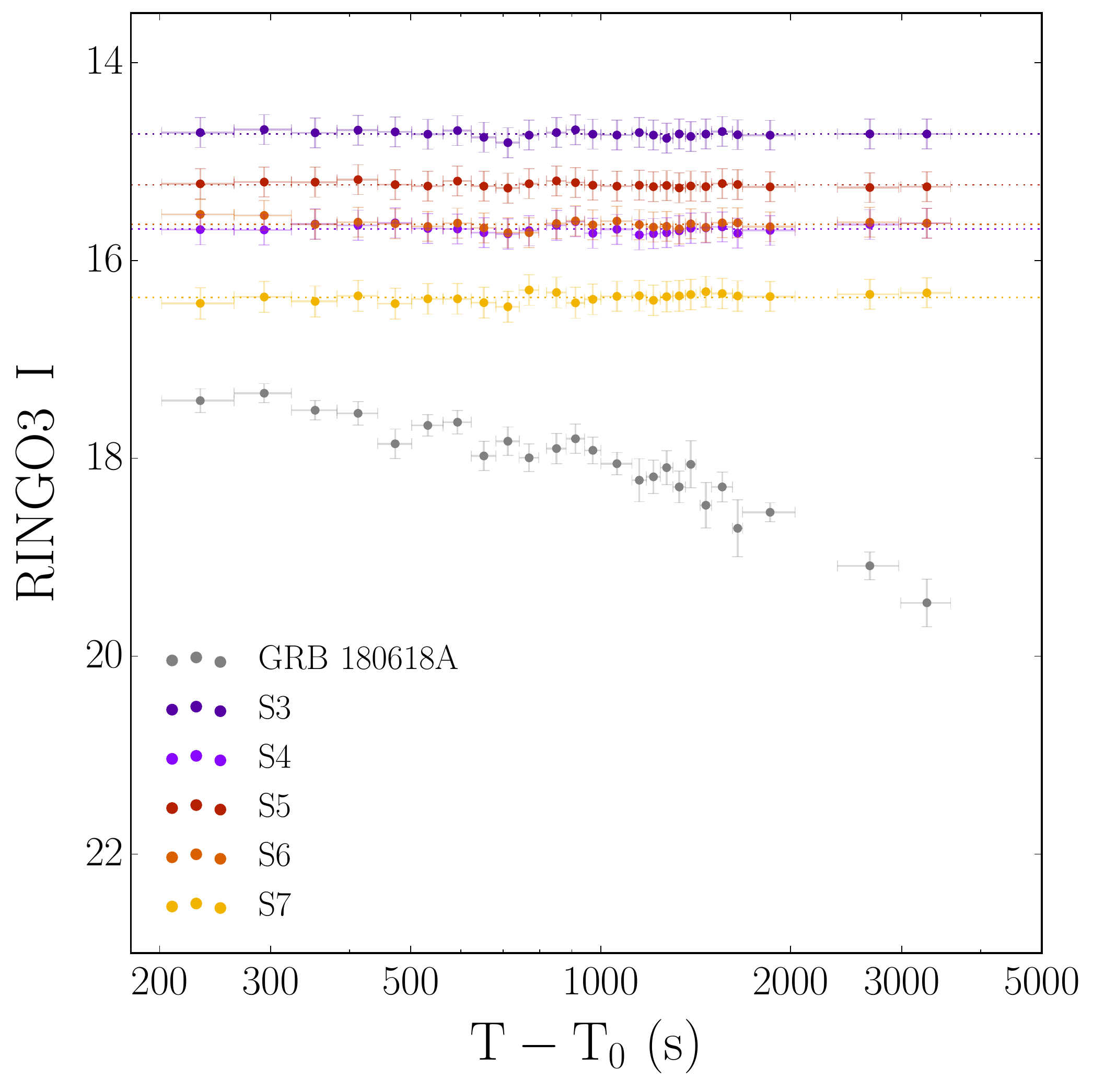}
\caption{The RINGO3 photometric analysis of Figure~\ref{fig:fov} field stars (S) using the temporal binning of the GRB 180618A light curves. Note that the stars S1 and S2 are not included in the analysis because the LT repointed to the GRB 180618A coordinates after the IO:O observations, and they fall outside of the revised RINGO3 field of view. From left to right, panels correspond to the RINGO3 {\it BV}, {\it R}, and {\it I} bands.}
\label{fig:starcheck}
\end{figure*}

\begin{figure*}[ht!]
\centering
\includegraphics[width=160mm]{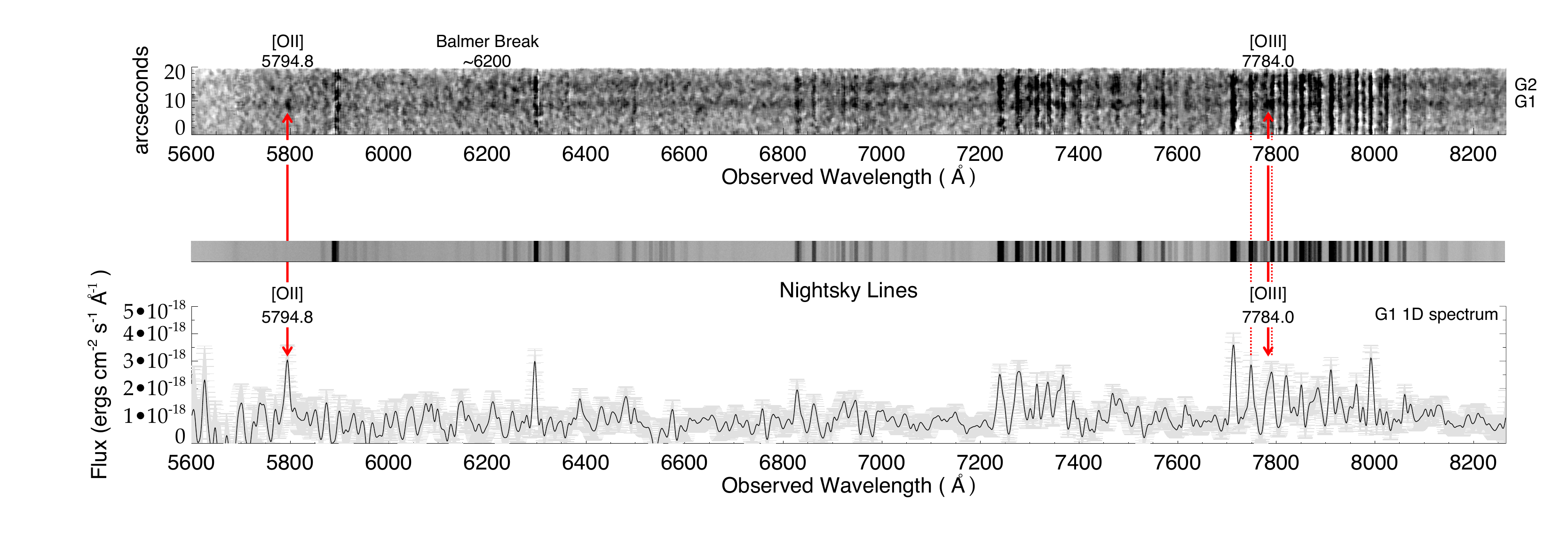}
\caption{From top to bottom, the two-dimensional MODS-2 spectra of the galaxy G2 and the GRB 180618A host galaxy (G1), the corresponding night-sky emission lines, and the one-dimensional spectrum of the G1 galaxy. In the G1 spectrum, we identify two oxygen emission lines (corresponding to [OII] and [OIII]) redshifted at $z=0.554\pm 0.001$. The night-sky spectrum displayed here demonstrates that both oxygen emission lines (marked in solid red arrows) are not unsubtracted night-sky emission lines; the [OII] line is not coincident with any night-sky line, and the [OIII] line is found between two (dotted red lines).}
\label{fig:GRB18_redshiftspec}
\end{figure*}

The UVOT photometry (Vega; see Table~\ref{tab:phot}) was derived from Level 1/2 products, which are already pre-processed by the telescope pipeline. We used the Level 1 event products to get higher temporal resolution. They were converted into sky-coordinated data with the {\it coordinator} tool from the {\it HEASoft} v6.22.1 package \citep{1995ASPC...77..367B}, and the hot pixels were removed with {\it uvotscreen}. The photometry was background-subtracted and measured with {\it uvotevtlc} using the default aperture radius of 5 arcsec from the instrument calibration \citep{2008MNRAS.383..627P,2011AIPC.1358..373B}. For the Level 2 images, we used the equivalent {\it uvotsource} tool. Furthermore, the images were aligned with {\it uvotskycorr} and co-added with {\it uvotisum}, requiring a minimum significance of $3\sigma$ for the detection of the optical transient. Those stacked frames in which the optical transient did not reach the signal-to-noise threshold are reported as $3\sigma$ flux upper limits in Table~\ref{tab:phot}.

The RINGO3 photometry (Vega; see Table~\ref{tab:phot}) was derived by integrating the source photon counts across the eight polaroid positions (e.g. \citealt{2020ApJ...892...97J}) ---thus cancelling any polarization signal. Each $10\,$minutes integration consists of ten single 1-minute exposures from which we individually derived the photometry using the {\it Astropy Photutils} package \citep{2016ascl.soft09011B}. If the signal-to-noise did not reach a minimum $3\sigma$ significance, we co-added consecutive frames. The RINGO3 magnitudes and flux density were absolute-calibrated in Vega system following a standard procedure (e.g., see \citealt{2020ApJ...892...97J}) with observations of five dereddened A0 type stars (BD +30 2355, BD +67 675, BD +25 2478, HD 96781, HD 208368; \citealt{2000A&A...355L..27H}). The standard stars and the GRB 180618A field observations were scheduled via the LT {\it phase2UI} using the same instrumental setup of the night of the burst, and they were executed on 13 and 16 June 2019. This spectral calibration added $\approx 0.07 \,$mag uncertainty to the photometry. To test for the instrument stability during observations, we checked the flux variability of five stars in the GRB 180618A field of view (see Figure~\ref{fig:starcheck}). Using the temporal binning of the GRB 180618A light curves, the stars displayed on average a $\approx 0.04 \,$mag deviation from the mean.

The IO:O camera photometry (AB; see Table~\ref{tab:phot}) was derived for each of the $10$-s individual frames. For the observing sequence at $7.4 \times 10^4 \,$s post-burst, we integrated the observations into a single $900\,$s exposure per band. The optical transient was not detected, and the flux upper limits are presented in Table~\ref{tab:phot} at $3\sigma$ confidence level. We calibrated the IO:O bands by cross-matching ten 12$-$17 mag stars from the Sloan Digital Sky Survey (SDSS) Data Release 12 catalogue \citep{2015ApJS..219...12A}.

\subsection{Optical Polarimetry} \label{sec:optpolreduc}

In the RINGO3 configuration, we measure the polarization of a source by extracting the flux at each of the eight polaroid positions (e.g. \citealt{2020ApJ...892...97J}), which we then convert to Stokes parameters ($q-u$) following \citealt{2002A&A...383..360C}. The polarization uncertainties are derived from a Monte Carlo error propagation, starting from $10^{6}$ flux values. 

We measured the instrumental $q-u$ using $\approx 75$ measurements of seven unpolarized standards per band (BD +32 3739, BD +33 2642, BD +28 4211, HD 212311, HD 14069, HD 109055, G191B2B; \citealt{1990AJ.....99.1243T,1992AJ....104.1563S}), which were observed during 200 days before and 10 days after the date of the burst. Note that we used an asymmetric time window of data given a small shift of $\Delta u \approx 0.005$ in the instrumental $u$ parameter after 10 days post-burst. For the chosen time window, there was no significant drift of the instrumental $q-u$ and the Pearson's correlation coefficients were low $\vert r \vert < 0.1$ with $p$-values$ > 0.3$.

To derive the most constraining polarization measurement for the GRB 180618A, we used the entire $10\,$minutes epoch corresponding to t$_1 = 203-800\,$s post-burst of the {\it BV} band, which is the RINGO3 band with the highest signal-to-noise. We detected the optical transient at a signal-to-noise $\approx 27$ in each of the eight images of the polaroid. At this signal-to-noise level, the observed polarization ($P \approx 1\%$) was within the instrument sensitivity, and we estimate a $2\sigma$ upper limit of $P_{BV} < 6.1\%$. Due to the slowly fading emission during the second and third observing epoch, at  t$_2 = 822-1417 \,$s and t$_3 = 1438-2035 \,$s post-burst respectively, the polarization upper limits could still be derived but were less well constrained, with $P_{BV, \, {\rm \lbrace t_2, t_3 \rbrace }} < 10.7 \%, 17.0 \%$ ($2\sigma$). For the {\it R} and {\it I} bands, the $2\sigma$ upper limits for the three epochs are $P_{R, \, {\rm \lbrace t_1, t_2, t_3\rbrace }} < 14.5 \%, 30.2 \%, 37.0 \%$ and $P_{I, \, {\rm \lbrace t_1, t_2, t_3\rbrace }} < 23.5 \%,  36.8 \%, 38.6 \%$. We note that the RINGO3 depolarization factor \citep{2021MNRAS.505.2662J} is negligible in the {\it BV} band (D$_{BV}=1$) and small in the {\it R} and {\it I} bands (D$_{\lbrace R, I \rbrace }=0.98, 0.94$).

\subsection{LBT Photometry} \label{sec:host_phot}

To search for the GRB 180618A host galaxy, and thus determine the GRB 180618A redshift, we used the Large Binocular Telescope (LBT; \citealt{2006SPIE.6267E..0YH}) ---an optical/infrared telescope with twin 8.4-m mirrors located on the Mt. Graham International Observatory, Arizona, USA.

On 5 April 2022, deep-field $r$- and $z$-band imaging of the GRB 180618A location were acquired with the Large Binocular Cameras (LBC; \citealt{2000SPIE.4008..439R,2008A&A...482..349G}). The total exposure time for each filter was 36 minutes and the data were reduced with the data reduction pipeline developed by the Istituto Nazionale di Astrofisica (INAF-Osservatorio Astronomico di Roma; \citealt{Fontana2014a}), which includes bias subtraction and flat-fielding, bad pixel and cosmic ray masking, astrometric calibration, and coaddition. The average seeing was $\approx 1.4 \,$arcsec and the mean airmass of the observations was $\approx 1.4$. The LBC photometry (AB) achieved limiting magnitudes of $r_{\rm lim} = 26.3\, $mag and $z_{\rm lim} = 25.6 \,$mag ($3\sigma$ limits). In both images, we found three galaxies (G1, G2, and G3; see Figure~\ref{fig:fov}) at a projected angular distance of $d_{\rm G1}=1.6 \,$arcsec, $d_{\rm G2}= 3.7 \,$arcsec, and $d_{\rm G3}= 7.7 \,$arcsec from the UVOT sub-arcsec localization of the GRB 180618A  \citep{2018GCN.22810....1S}. The G3 galaxy was already catalogued by the Sloan Digital Sky Survey (SDSS; \citealt{2015ApJS..219...12A}), with a photometric redshift of $z_{\rm G3}=0.54\pm 0.03$, and brightness $r_{\rm band, G3} = 22.4 \pm 0.3\,$mag and $z_{\rm band, G3}  = 21.2 \pm 0.6 \,$mag. The G1 and G2 galaxies were uncatalogued in SDSS, but identified in the \cite{2022arXiv220409059O} and \cite{2022arXiv220601763F} surveys of short GRBs host galaxies. From the LBC images, we measured $r_{\rm band, G1}=22.98\pm 0.06\,$mag, $r_{\rm band, G2}=23.58 \pm 0.11\,$mag in the $r$ band, and $z_{\rm band, G1} =22.62 \pm 0.10\,$mag,  $z_{\rm band, G2} =22.48 \pm 0.09\,$mag in the $z$ band.

\subsection{LBT Spectroscopy} \label{sec:host_spec}

On 8 April 2022, optical spectroscopy of the G1 and G2 galaxies was obtained with the Multi-Object Double Spectrographs (MODS; \citealt{2010SPIE.7735E..0AP}), i.e. MODS-1 and MODS-2. Each MODS contains a red and blue channel for spectroscopy. Both MODS were configured to use the Dual Grating mode ($0.32\,\mu\textnormal{m}-1.05\,\mu$m coverage) and a 1.2 arcsec wide slit ($R \approx 630-1350$ resolution). A position angle $179.3^{\circ}$ was used for observations. The position angle was selected so that the target galaxies and the foreground star ($\approx$47 arcsec away) were vertically aligned. The foreground star was aligned in the centre of the slit as the target galaxies were too faint to be detected in the acquisition images without incurring too large of an overhead. Two exposures of $20 \,$minutes each were obtained in each channel, with each MODS. The mean airmass of the observations was $\approx 1.85$. The mean seeing (as measured from the off-axis wavefront sensor and guiders) was 1.92$\pm$0.05 arcsec and 1.84$\pm$0.04 arcsec for MODS-1 and MODS-2, respectively. Observations of the spectrophotometric star BD+33 2642 were used to flux calibrate the data and remove the instrumental signatures from the data.

The MODS data were reduced first with the {\it modsCCDRed} v2.04 package developed by the MODS team \citep{2019zndo...2647501P} to remove the bias and flat-field the data using a slitless pixel flat. Next, custom IRAF scripts \citep{1986SPIE..627..733T} were used to extract along the central slit using a stellar trace. The observations of the spectrophotometric standard were combined to measure the trace of the dispersion along the entire slit. This trace was used along with the wavelength calibration from arc-lamp lines to rectify the tilt in both x and y directions for the full frame (8192 pixels $\times$ 3072 pixels). This step made the x-axis parallel to the dispersion direction and the y-axis purely parallel with the spatial extent along the slit. Final wavelength calibration was cross-checked with known strong auroral skylines in the blue ([OI] $\lambda = 5577.3 \,$\AA) and red ([OI] $\lambda=6300.3 \,$\AA) channels. One-dimensional spectra were then extracted from each channel using a $1.85 \,$arcsec wide aperture. This value was chosen to match the mean seeing of the observations, maximizing the signal-to-noise. Next, the spectra were flux-calibrated using the spectro-photometric standard star. Telluric features were removed from the red channels using a normalized spectro-photometric standard spectrum.

Inspection of the data showed no flux from the galaxies in the blue channels of both MODS. A faint but significant continuum was detected for the galaxies in MODS-2 (see Figure~\ref{fig:GRB18_redshiftspec}), but not in MODS-1. The absence of flux in MODS-1 is consistent with a known technical issue\footnote{\url{https://scienceops.lbto.org/mods/preparing-to-observe/sensitivity/}} in which the sensitivity of the instrument has decreased by a factor of 1.6 since the 2011 commissioning. As a sanity check, the acquisition images were re-checked to confirm no differences in the alignment of the foreground star. The continua of the target galaxies are clearly visible in the individual background-subtracted exposures for the red channel of MODS-2 (see Figure~\ref{fig:GRB18_redshiftspec}). The spatial position (y-direction) of the continua from the target galaxies detected in the MODS-2 red channel exactly matched the angular separation between them and the foreground alignment star ---as measured from the LBC images (see Figure~\ref{fig:fov}). The final calibrated MODS-2 red channel data have a fixed instrumental resolution of 8.19 \AA \, per resolution element covering $5600-10100 \,$\AA, which corresponds to a resolution of $R \approx 630-1350$ and 0.84 \AA/pixel. These values were confirmed using the arc lamps.

%%%%%%%%%%%%%%%%%%%%%%%%%%%%%%%%%%%%%%%%%%%%%%%%%%%%%%%%%%%%%%%%%%%%%%%%%%%%%%%%

\section{Results} \label{sec:results}

In Section \ref{sec:opt}, we present the temporal and spectral analysis of the optical emission. In Section \ref{sec:redhost}, we estimate the redshift of the GRB 180618A using the ultraviolet and optical photometry of the transient emission, and we then associate the GRB 180618A with its host galaxy. In Sections \ref{sec:gamma_durprop}-\ref{sec:energetics}, we study the gamma-ray properties of the short GRB 180618A and its extended gamma-ray emission.

\subsection{Optical Emission} \label{sec:opt}

\begin{figure}[ht!]
\centering
\includegraphics[width=\columnwidth]{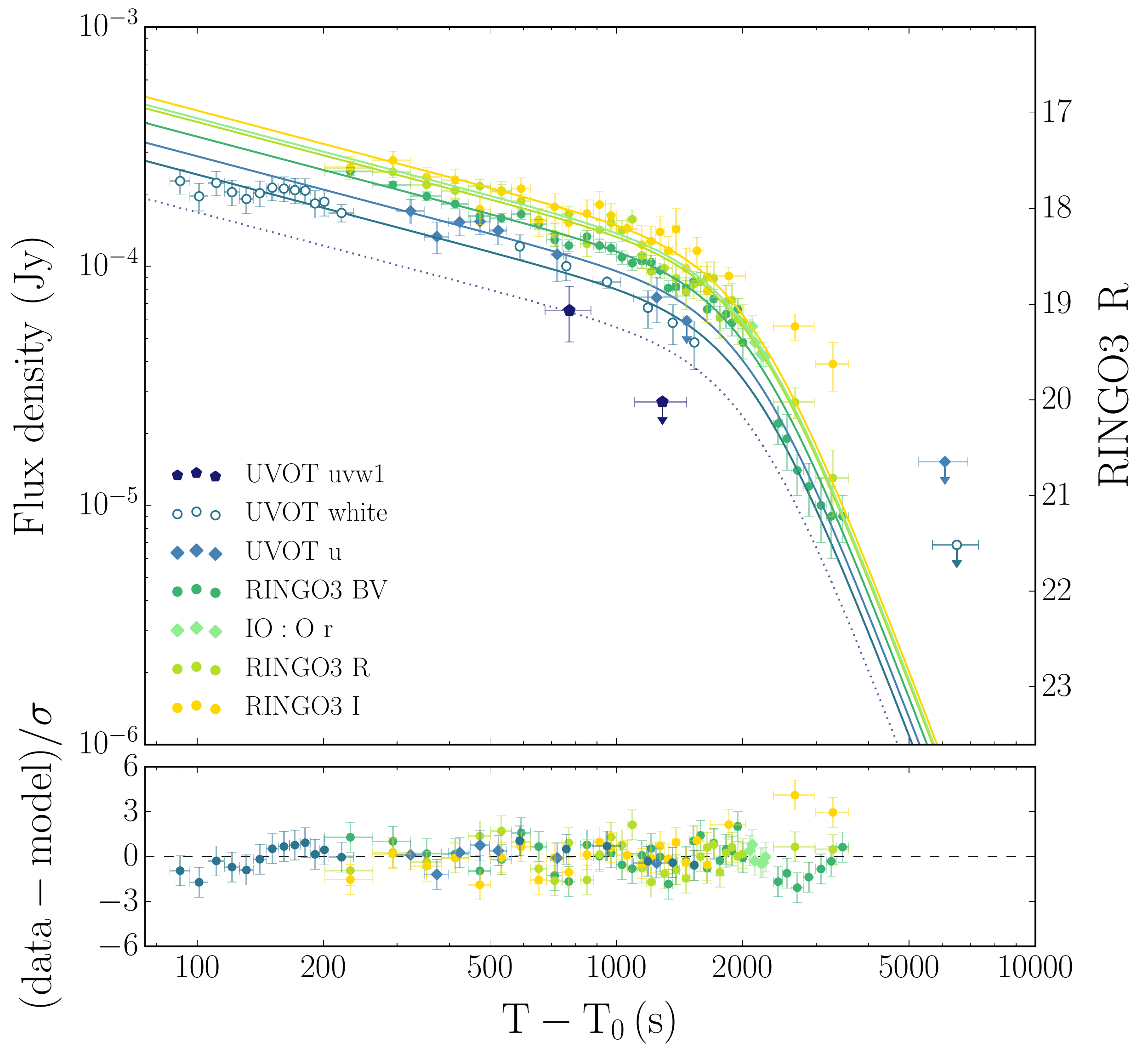}
\caption{The RINGO3 {\it BV,R,I} bands and the UVOT {\it white, u} bands data modelled with smoothly-connected broken power laws that have a common time break across bands. The results of the fit are a break at $t_{\rm break} =  2120 \pm 60 \,$s post-burst, power-law indexes $\alpha_1 = 0.46 \pm 0.02$ pre-break and $\alpha_2 =  4.6 \pm 0.3$ post-break with $\chi^2/$dof$ = 128/116$. Also, we show with a dotted line the best-fitting model normalized to the near-ultraviolet UVOT {\it uvw1} band, which has a power-law decay $\alpha_{uvw1}>1.7$, pre-break. Detections have $1\sigma$ error bars, and non-detections are presented as $3\sigma$ upper limits.}
\label{fig:LC_180618A_optical}
\end{figure}

We simultaneously fitted the UVOT {\it white}, RINGO3 {\it BV,R,I} and IO:O {\it r} optical light curves with smoothly-connected broken power laws \citep{1999A&A...352L..26B,2007AandA...469L..13M}, i.e. \linebreak $F =F_0  [(t/t_{\rm break})^{n \alpha_{1}} + (t/t_{\rm break})^{n \alpha_{2}}] ^{-1/n}$, fixing the time break across bands and the smoothness parameter to $n=1$ for convergence \citep{2008A&A...491..183P}. This serves to help us understand the overall decay rate of the emission, as well as to check for colour evolution in the residuals of the best-fitting model. The emission initially decays with $\alpha_{\rm opt, 1} = 0.46 \pm 0.02$ and suffers a sharp break at $t_{\rm break} = 2120 \pm 60 \,$s post-burst, with $\alpha_{\rm opt, 2} = 4.6 \pm 0.3$ (see Figure~\ref{fig:LC_180618A_optical}).

\begin{figure}[ht!]
\centering
\includegraphics[width=\columnwidth]{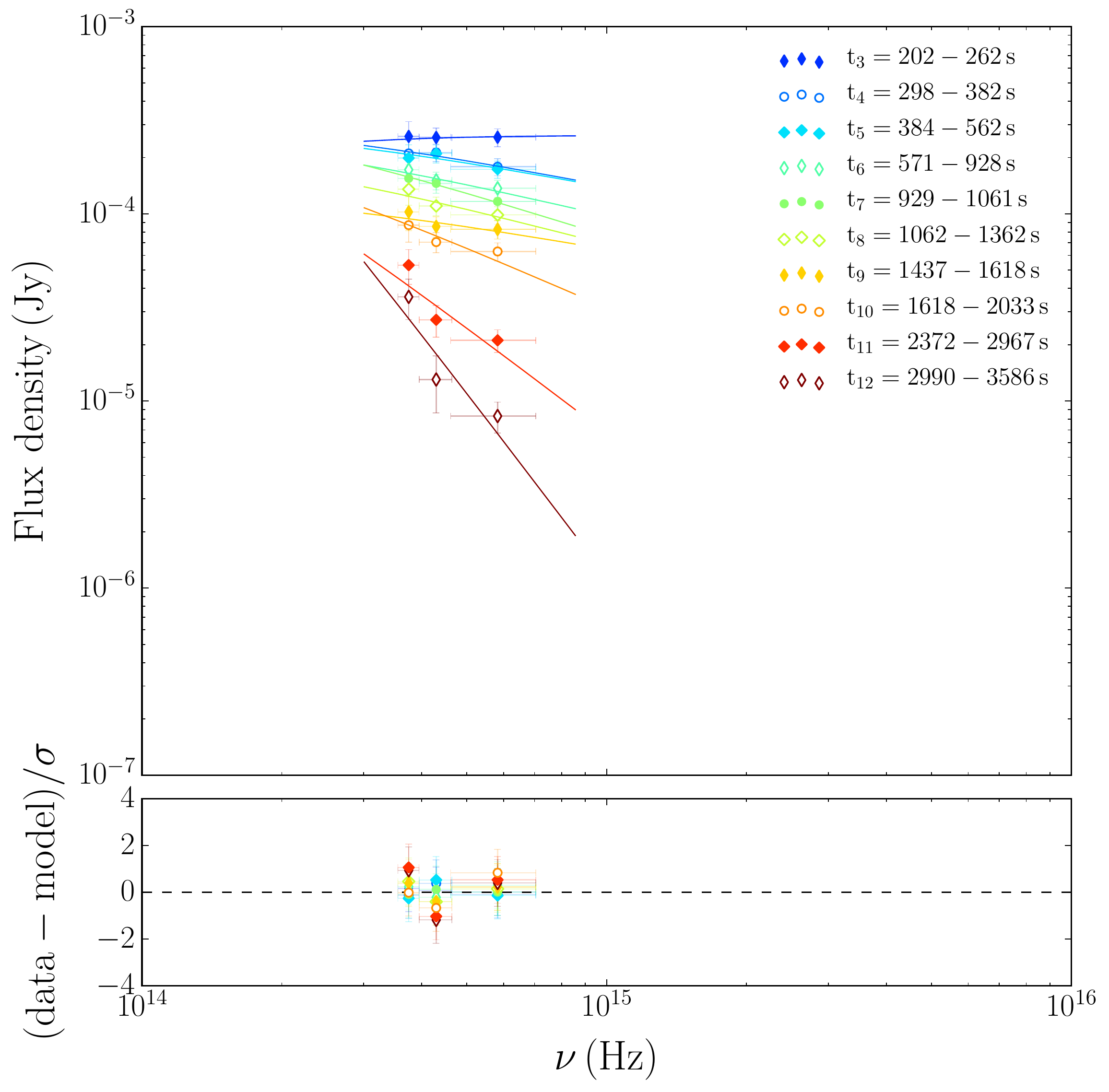}
\caption{Modelling of the GRB 180618A RINGO3 optical emission with power-law models. Note that the model accounts for Galactic dust extinction, but does not include the host galaxy dust contribution.}
\label{fig:PLfit}
\end{figure}

This extreme flat-to-steep decay evolution is rare in GRB light curves. To our knowledge, the only other GRB that has shown a similar flat-to-steep decay rate transition ($\alpha_{\rm opt, 1}=0.44 ^{+0.08} _{-0.21}$ to $\alpha_{\rm opt, 2}=5.3 \pm 0.2$) was found in the fainter optical emission of the short GRB 070707 \citep{2008A&A...491..183P} and at a much later time at  ${\approx 1.8\,}$days post-burst. In addition, in GRB 180618A the spectral evolution across the break is chromatic; the RINGO3 {\it I} band emission is significantly underestimated by $>3\sigma$ after the break, and the normalized best-fitting model overestimates the $3\sigma$ photometric upper limit of the near-ultraviolet UVOT {\it uvw1} band. Furthermore, we individually fitted the flux decay rate of the RINGO3 {\it BV, R, I} and IO:O {\it r} bands after the break with a power law, finding moderate slopes for redder bands, i.e. $\alpha_{{\rm opt}, \lbrace BV, r, R, I \rbrace}= 4.6 \pm 0.5, 3.5 \pm 0.5, 3.2 \pm  0.1, 1.4 \pm 0.2$.

Using {\it Xspec} v12.9.1 \citep{1996ASPC..101...17A} and $\chi^2$ statistics, we modelled the RINGO3 data with a power law that accounts for the Milky Way dust extinction, i.e. E(${B-V}$)$_{\rm MW} = 0.065 \pm 0.003$ \citep{1998ApJ...500..525S}. We find that the optical photon index significantly evolves during observations, from $\beta_{\rm opt, PI}= 0.7 \pm 0.4$ at $t=202 - 263\,$s post-burst to $\beta_{\rm opt, PI} = 4.0 \pm 0.8$ at $t=2990-3586\,$s post-burst; see results in Figure~\ref{fig:LC_GRB180618A}-bottom panel, Figure~\ref{fig:PLfit} and Table~\ref{tab:PL}. Note that if we include the UVOT data, we need to add the host galaxy dust extinction to the model, which shifts all the optical photon indexes by the same amount towards harder values (i.e. the relative evolution of the $\beta_{\rm opt, PI}$ remains the same).

\subsection{Redshift and Host Galaxy} \label{sec:redhost}

\begin{figure*}[ht!]
\centering
\plottwo{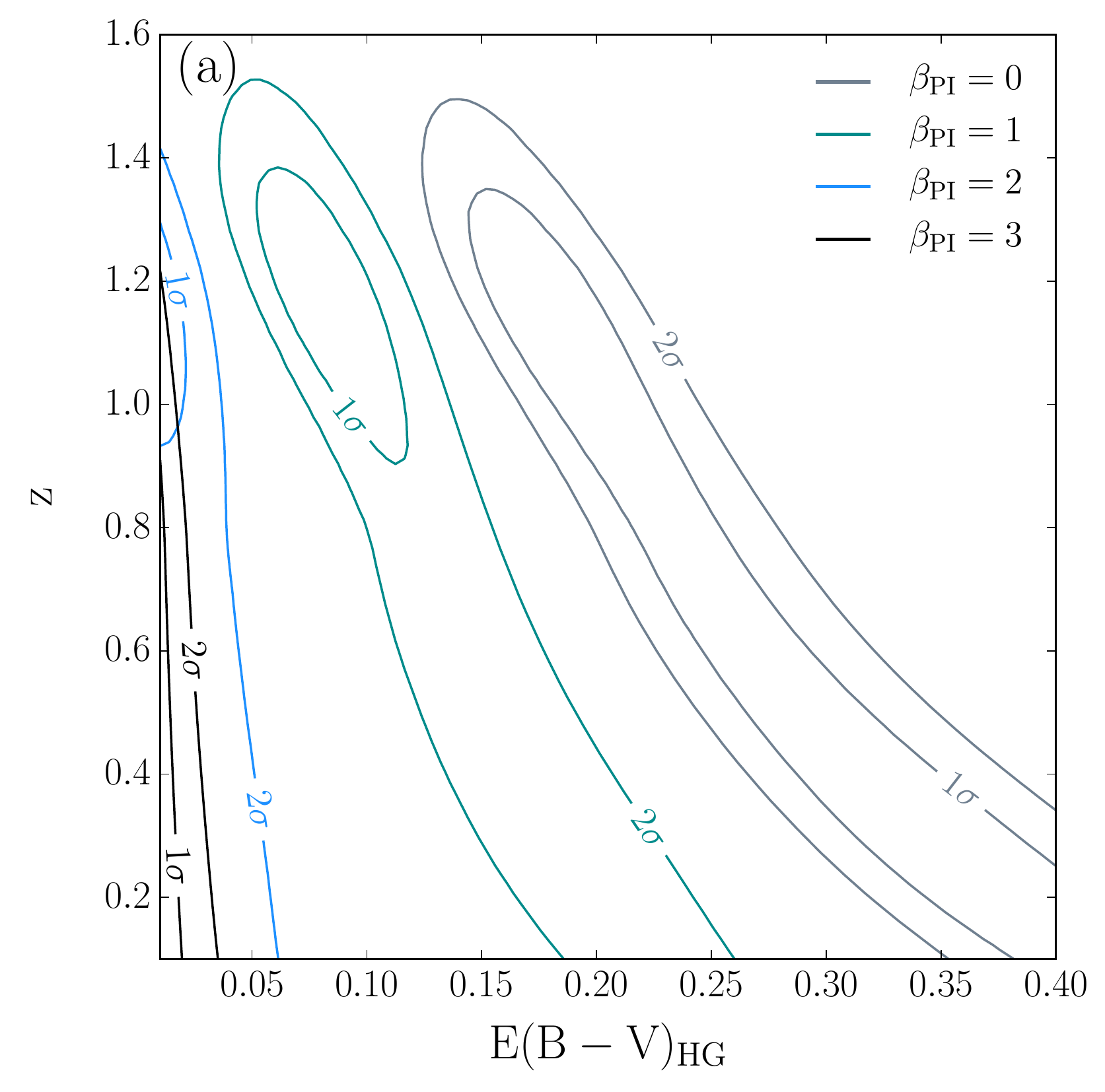}{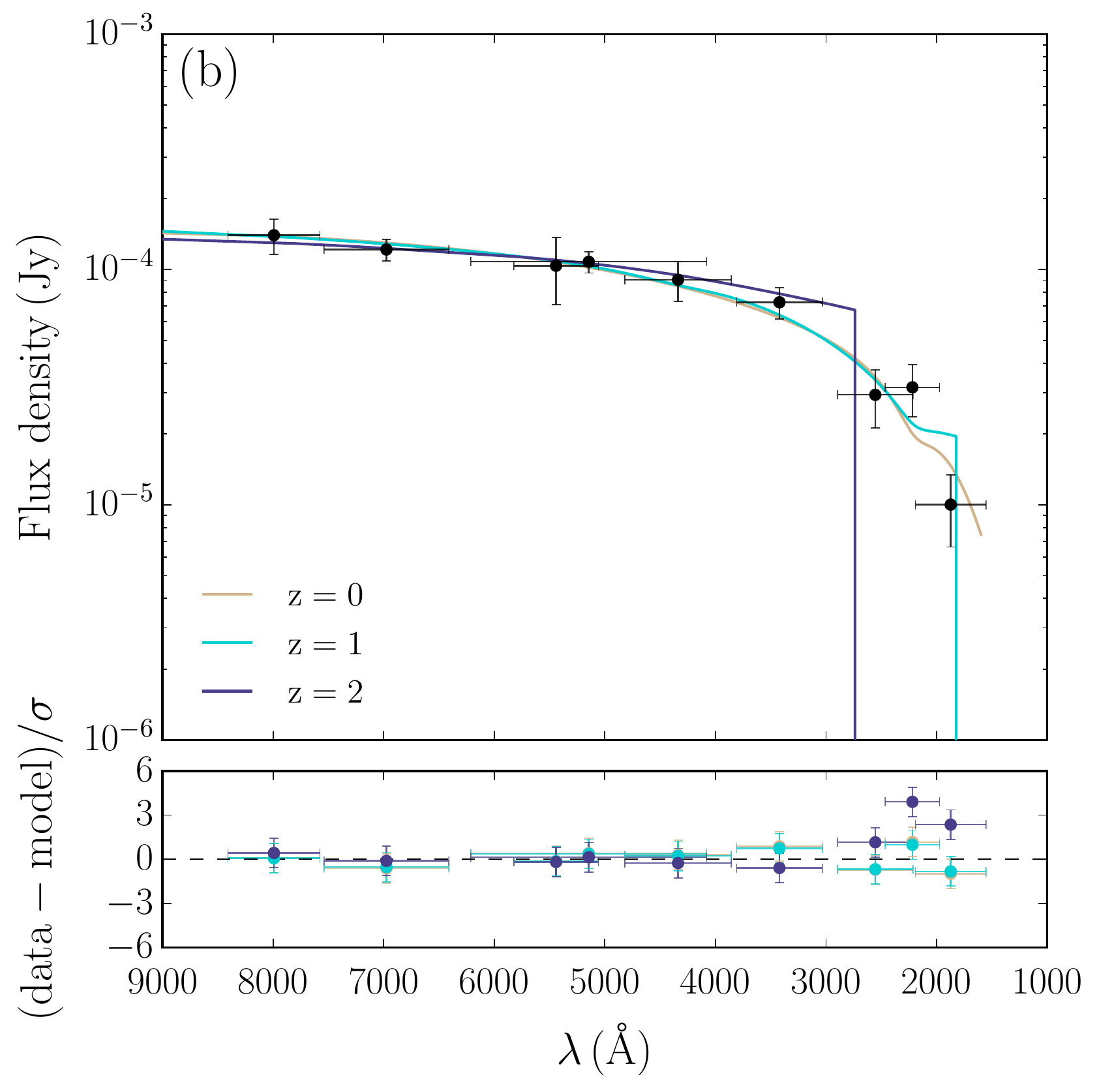}
\caption{Estimation of the GRB 180618A redshift from the cotemporal GRB 180618A UVOT and RINGO3 data modelled with a dust-absorbed power law that includes the redshifted Lyman-limit break. (a) The $\chi^2$ distribution of the best-fitting redshifts for different values of the host galaxy extinction E($B-V$)$_{\rm HG}$ and photon index $\beta_{\rm PI}$. The confidence level contours at $2\sigma$ level indicate a redshift $z<1.5$ for the GRB 180618A. (b) Best-fitting models for redshifts $z=0,1,2$. }
\label{fig:GRB18_redshift}
\end{figure*}

We rule out a high redshift origin given the detection of the GRB 180618A optical counterpart with all the UVOT filters \citep{2011A&A...526A.153K}. We used the {\it uvot2pha} tool to convert the UVOT data to {\it Xspec} spectral files, and a dust-absorbed power law that includes both Milky Way and host galaxy contributions, i.e. E($B-V$)$_{\rm HG}$. Taking into account the spectral coverage up to the far-ultraviolet of the UVOT {\it uvw2} filter and using the redshifted Lyman-limit break at $\lambda_{\rm obs} =912(1+z) \,$\AA , we fitted the co-temporal UVOT and RINGO3 data corresponding to $550-1600\,$s post-burst using $\chi ^2$ statistics. We estimate a redshift $z < 1.5$ at $2\sigma$ confidence level from the E($B-V$)$_{\rm HG}-\beta_{\rm PI}-z$ parameter space (see Figure~\ref{fig:GRB18_redshift}-a). Note that not knowing the GRB 180618A intrinsic spectral slope does not affect the redshift constraint but adds large uncertainty in determining the host galaxy dust contribution. See also the best-fitting models for the redshifts $z=0, 1, 2$ in Figure~\ref{fig:GRB18_redshift}-b.

Among all three candidate galaxies detected in the LBT images, G1 is the most likely host galaxy of GRB 180618A given its proximity (see Figure~\ref{fig:fov}). In G1 galaxy spectrum (see Figure~\ref{fig:GRB18_redshiftspec}), we detect two emission lines at $\lambda_{\rm obs, OII}=5794.8\,$\AA \, and $\lambda_{\rm obs, OIII}=7784.0\,$\AA, corresponding to the unresolved [OII] $\lambda= 3726-3729 \,$\AA \, doublet and the [OIII] $\lambda= 5007 \,$\AA \, line, respectively. That implies a spectroscopic redshift $z_{\rm G1} =0.554 \pm 0.001$ for the G1 galaxy. This value is consistent with \cite{2022arXiv220409059O} and \cite{2022arXiv220601764N} estimates of photometric redshifts $z=0.4^{+0.2} _{-0.1}$ and $z=0.52^{+0.09} _{-0.11}$, respectively. We find that the probability of an accidental alignment \citep{2002AJ....123.1111B,2010ApJ...722.1946B,2013ApJ...776...18F} of the GRB 180618A and G1 galaxy is low, with $p_{\rm d} \approx 0.02$. Therefore, we associate the G1 redshift ($z=0.554$) with the GRB 180618A. For the G2 galaxy spectrum, we do not detect emission lines. However, there is a clear drop of the continuum flux bluewards of $\lambda_{\rm obs} \approx 6200 \,$\AA \, that we identify as the $\lambda=4000\,$\AA \, break, corresponding to a redshift $z_{\rm G2} \approx 0.55$. Therefore, the LBT spectroscopic analysis suggests that all three galaxies (G1, G2, and G3) are at a similar redshift. 

The GRB 180618A lies in the outskirts of its host, at $10 \,$kpc from the centre of the galaxy (see also \citealt{2022arXiv220409059O,2022arXiv220601763F}). This is consistent with the large offsets found in short GRBs, and in disagreement with that of long GRBs \citep{2010ApJ...722.1946B,2013ApJ...776...18F,2014ApJ...792..123B,2022arXiv220601763F}. Note that short GRBs are usually found with offsets to star-forming disk galaxies or even further away from their elliptical host \citep{2013ApJ...776...18F,2014ApJ...792..123B}. Similar to the environment of the GRB 180168A, \cite{2013ApJ...776...18F} found that about $30\%-45\%$ of short GRBs happen where there is no optical light, i.e. negligible stellar mass. Like GRB 180618A, most short GRBs display signs of migration from their birth sites, likely due to natal kicks in binaries \citep{2003MNRAS.345.1077R,2010ApJ...725L..91K,2013ApJ...776...18F}. This is consistent with short GRBs exploding in low ambient density, thus producing fainter afterglows \citep{2010ApJ...722.1946B,2011ApJ...734...96K,2015ApJ...815..102F}.

\subsection{Duration of the Prompt Gamma-ray Emission} \label{sec:gamma_durprop}

\begin{figure*}[ht!]
\centering
\includegraphics[width=\textwidth]{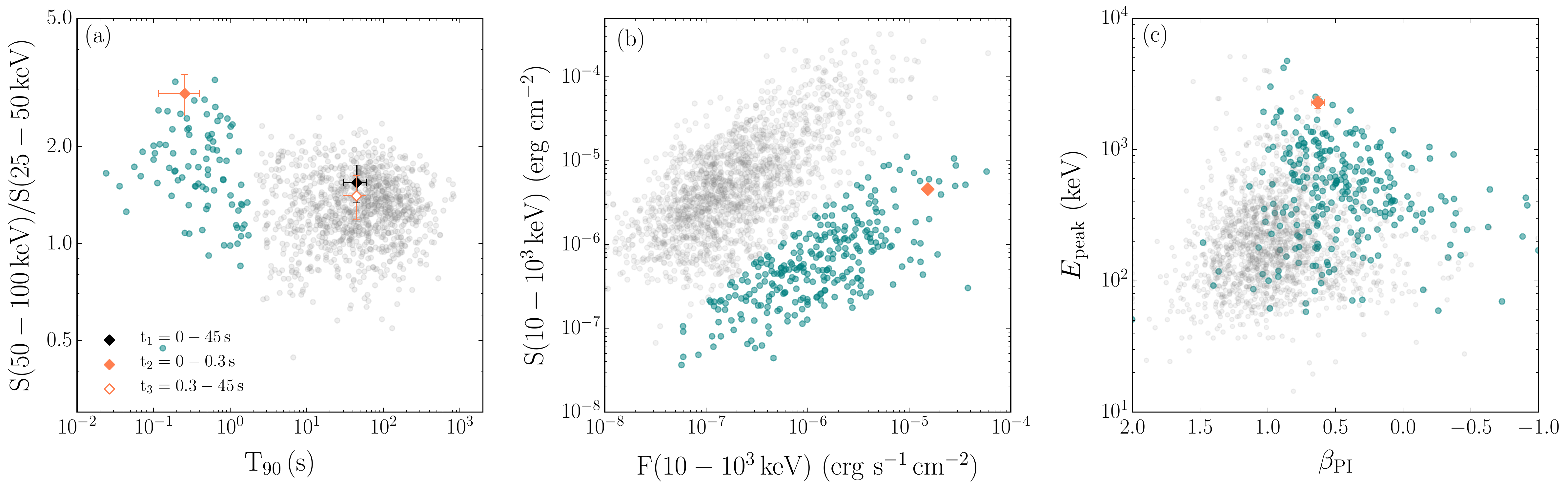}
\caption{High-energy properties of the GRB 180618A and comparison with catalogue GRBs; those with duration $T_{\rm 90}>2 \, $s are displayed in grey, and those with $T_{\rm 90}<2 \,$s in blue \citep{1993ApJ...413L.101K}. (a) The BAT band hardness ratio and duration of the GRB 180618A ($t_1$), the $\approx 0.3 \, $s short gamma-ray pulse ($t_2$), and its $\approx 45 \, $s extended gamma-ray emission ($t_3$). The background data corresponds to the short-hard and long-soft bimodal clustering of the BAT GRB catalogue \citep{1993ApJ...413L.101K,2016ApJ...829....7L}. (b) Comparison of the  $\approx 0.3 \, $s short GRB 180618A flux F and the fluence S (marked in orange) with the GRB sample from the GBM catalogue \citep{2020ApJ...893...46V,2021ApJ...913...60P}. Note that the GRB 180618A flux and fluence have been recalculated to match the 10$-$10$^3$ keV energy range of the GBM catalogue. (c) The photon index $\beta_{\rm PI}$ and the peak energy $E_{\rm peak}$ of the GRB 180618A (marked in orange).}
\label{fig:LC_180618A_hr}
\end{figure*}

We calculated the duration ($T_{90}$) of GRB 180618A \citep{2016ApJ...829....7L,2022ApJ...927..157M}, corresponding to the time interval in which $90\%$ of the burst fluence is released, using the 64-ms binned and background-subtracted GRB 180618A light curves of the BAT. Using the $battblocks$ tool \citep{1998ApJ...504..405S}, a {\it FTOOLS} released as part of {\it HEASoft} \citep{1995ASPC...77..367B}, we find $T_{90} = 45  \pm 10 \,$s for the low-energy spectral range of the BAT (i.e. 15$-$100 keV), and $T_{90}=0.26  \pm 0.14\,$s at 100$-$350 keV. The duration of GRB 180618A at the low-energy bands is two orders of magnitude higher than at high-energy bands, confirming two spectral components: a short-hard GRB and extended soft gamma-ray emission below the 100 keV energy band \citep{2018GCN.22794....1H,2018GCN.22796....1S,2018GCN.22822....1S}. We find that the 0.3-s short-duration GRB 180618A belongs to the hardness-duration cluster of short GRBs, and that it is one of the hardest detected by the BAT (within the top $\approx 0.5\%$; see Figure~\ref{fig:LC_180618A_hr}-a). Note also that \cite{2018GCN.22796....1S} found negligible spectral lag for the short-duration gamma-ray pulse of GRB 180618A ---a spectral property typical of short-hard GRBs.

Furthermore, GRB 180618A is a classically short GRB in terms of the duration reported in the GBM GRB catalogue \citep{2020ApJ...893...46V}, with $T_{\rm 90} (50-300 \, {\rm keV})=3.7\pm0.6\,$s. For the GBM, \cite{2020ApJ...893...46V} found that the threshold separating short and long GRBs is $T_{\rm 90} = 4.2 \,$s ---instead of the $T_{\rm 90} \approx 2 \, $s of the Burst And Transient Source Experiment (BATSE; \citealt{1993ApJ...413L.101K}).

\subsection{Spectral Properties of the Prompt Gamma-ray Emission} \label{sec:gamma_specprop}

We performed a Swift-BAT and Fermi-GBM joint spectral fit in {\it RMFit}\footnote{\url{https://fermi.gsfc.nasa.gov/ssc/data/analysis/rmfit/}} v4.3.2. We used the GBM time-tagged events (TTE) data from the NaI 3/4 and BGO 0/1 detectors, from which we respectively selected the 8$-$900 keV and 200 keV$-$40 MeV spectral regimes. The 15$-$150 keV BAT spectra were extracted with the $batbinevt$ tool. Using C-statistics, we modelled the $\nu F_{\nu}$ spectrum of the $\approx 0.3\,$s short GRB with a simple power law. A power law model overestimates the data over $\approx 1 \,$MeV; consequently, we modelled the high-energy break with a cutoff power-law model (e.g., see \citealt{2021ApJ...913...60P}). The best-fitting model suggests that the $\nu F_{\nu}$ spectrum has a hard slope with mean photon index $\beta_{\rm PI} = - \alpha_{\rm CL} = 0.63 \pm 0.05$,  peak energy $E_{\rm peak}= 2290 \pm 238 \, $keV, and fluence $S(10-10^4 \, {\rm keV})= (5.6 \pm 0.4) \times 10^{-6} \, $erg cm$^{-2}$. The photon index is average compared to other short-hard GRBs detected by the GBM \citep{2020ApJ...893...46V,2021ApJ...913...60P}. Yet, the GRB 180618A is one of the hardest and most energetic gamma-ray pulses among short GRBs; it is within the $\approx 1\%$ percentile in terms of high-frequency peak energy and within the top $\approx 5\%$ in total energy released (see Figs.~\ref{fig:LC_180618A_hr}-b,c).

For the spectrum corresponding to the $\approx 0.3-45 \,$s extended gamma-ray emission, the best-fitting photon index of a power-law model is $\beta_{\rm PI} =1.51 \pm 0.09$. This intermediate photon index suggests that the spectrum has a cutoff (e.g., \citealt{2021ApJ...913...60P}). In order to constrain the peak energy, we fixed the low-energy index of the cutoff power-law model to the average of the GBM catalogue \citep{2021ApJ...913...60P} of short ($\alpha_{\rm CL}=-0.6$) and long GRBs ($\alpha_{\rm CL}=-1$). We find $E_{\rm peak} = 87\pm 18 \, {\rm keV} $ and $S(10-10^4 \, {\rm keV})= (6 \pm 1) \times 10^{-7} \, $erg cm$^{-2}$ for $\alpha_{\rm CL}=-0.6$, and $E_{\rm peak} = 125\pm 45 \, {\rm keV} $ and $S(10-10^4 \, {\rm keV})= (8 \pm 2) \times 10^{-7} \, $erg cm$^{-2}$ for $\alpha_{\rm CL}=-1$, consistent with \cite{2018GCN.22822....1S} findings.

To determine the temporal evolution of the photon index and the peak energy, we fitted a cutoff power-law model to the 8 keV$-$40 MeV GBM spectrum of each light curve bin. To avoid merging peaks and valleys, we used a constant binning as opposed to a fixed signal-to-noise (e.g., \citealt{2010ApJ...725..225G}). Using the default $64$-ms binned light curves, we find that in a time span of less than $260 \,$ms, the peak energy increases from $E_{\rm peak} = 427 \pm 138 \, $keV to $E_{\rm peak} = 2593 \pm 473\, $keV with constant photon index. As seen in other spectroscopically-resolved short GRBs \citep{2010ApJ...725..225G}, the peak energy tracks the light curve with a strong soft-hard-soft spectral evolution in a short time period; see the peak energy evolution in Figure~\ref{fig:LC_180618A_gamma}-b and the photon index in Figure~\ref{fig:LC_GRB180618A}-bottom panel.

\subsection{Intrinsic Energetics} \label{sec:energetics}

\begin{figure}[ht!]
\centering
\includegraphics[width=\columnwidth]{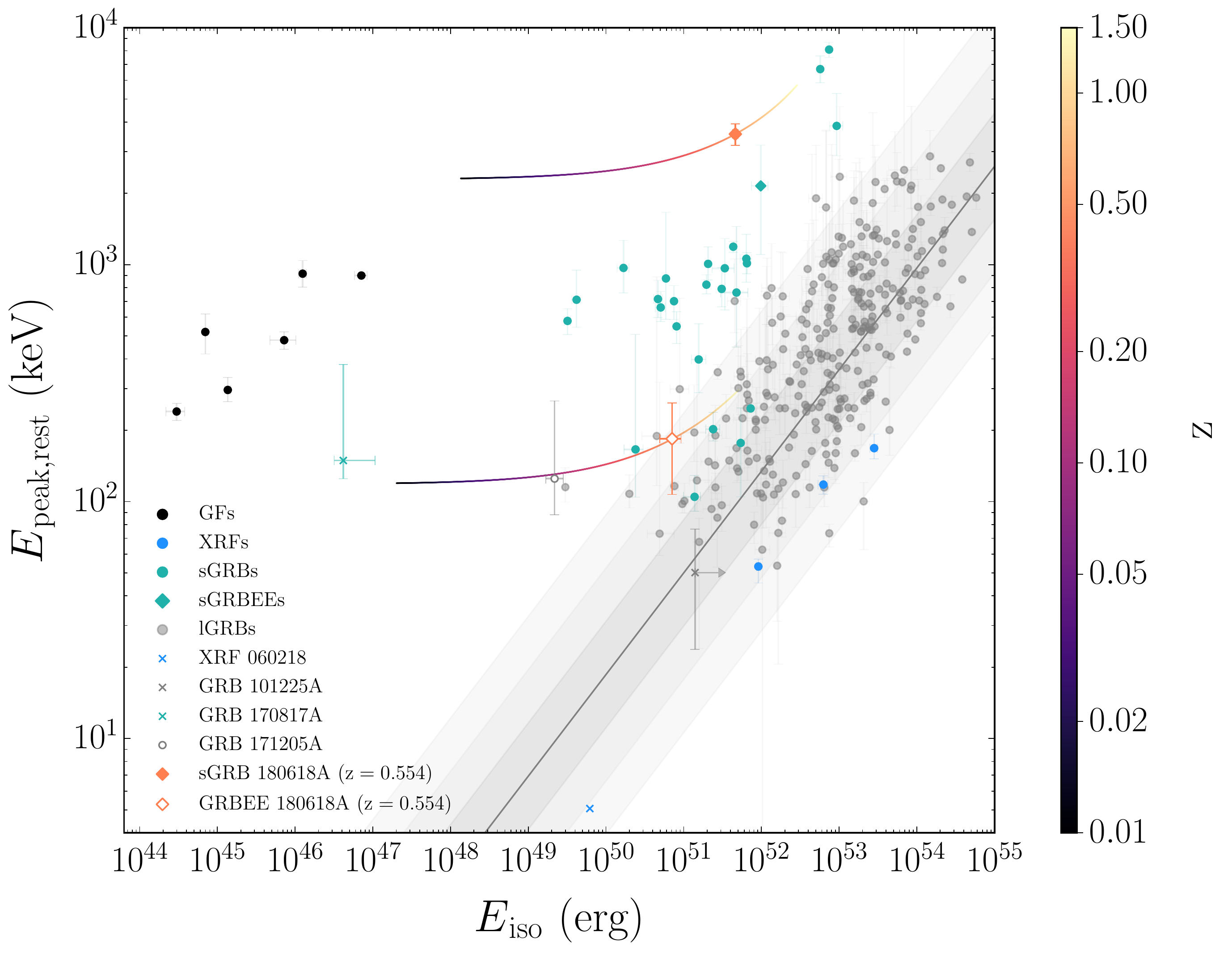}
\caption{Rest-frame peak energy ($E_{\rm peak}$) and isotropic-equivalent energy ($E_{\rm iso}$) of the short GRB 180618A and its extended gamma-ray emission at redshift $z=0.554$. The GRB 180618A high-energy parameters are also displayed for the redshift range $z=0.01-1.5$. For reference, we include those GRBs with known redshifts classified as giant flares from magnetars (GFs; \citealt{2020ApJ...903L..32Z}), short GRBs (sGRBs; \citealt{2014MNRAS.442.2342D,2015MNRAS.448..403C, 2017ApJ...850..161T,2021ApJ...908...83T}), short GRBs with extended gamma-ray emission (sGRBEEs; \citealt{2014MNRAS.442.2342D}), long GRBs (lGRBs; \citealt{2017ApJ...850..161T,2021ApJ...908...83T}), and X-ray flashes (XRFs; \citealt{2021ApJ...908...83T}). For the population of long GRBs, we represent the best-fitting line from \cite{2021ApJ...908...83T}, and the corresponding $1 \sigma$, $2 \sigma$, $3 \sigma$ confidence level. We also add the GRB counterpart of the gravitational wave event GW170817 \citep{2018NatCo...9..447Z}, and three GRBs with early afterglows interpreted as black body emission from either a thermalized jet or cocoon: XRF 060218 \citep{2006Natur.442.1008C,2011Natur.480...72T}, GRB 101225A \citep{2011Natur.480...72T}, and GRB 171205A \citep{2018A&A...619A..66D,2019Natur.565..324I}.}
\label{fig:180618A_class}
\end{figure}

Short GRBs have typically lower fluences and thus follow a different peak energy ($E_{\rm peak}$) and isotropic energy ($E_{\rm iso}$) relation than long GRBs \citep{2006MNRAS.372..233A,2009A&A...496..585G}. If we introduce the k-correction \citep{2001AJ....121.2879B} to the 1$-$10$^4$ keV rest-frame energy band for redshift $z=0.554$, we obtain an isotropic-equivalent energy $E_{\rm iso}= (4.6 \pm 0.4) \times 10^{51} \, $erg and luminosity $L_{\rm iso}= (1.9 \pm 0.2) \times 10^{52} \, $erg s$^{-1}$ for the short GRB 180618A, and $E_{\rm iso}= (7 \pm 2) \times 10^{50} \, $erg and $L_{\rm iso}= (2.4 \pm 0.7) \times 10^{49} \, $erg s$^{-1}$ for the extended gamma-ray emission. See also the high-energy properties for redshifts $z=0.01-1.5$ in Figure~\ref{fig:180618A_class}. For redshifts $z \gtrsim 0.1$, the GRB 180618A lays within the cluster of short GRBs ---it is one of the hardest and not compatible with the long GRB population. Overall, the gamma-ray properties and the GRB-host galaxy offset confirm the merger nature of the GRB 180618A (e.g., see \citealt{2009ApJ...703.1696Z,2022arXiv220410864R}).

%%%%%%%%%%%%%%%%%%%%%%%%%%%%%%%%%%%%%%%%%%%%%%%%%%%%%%%%%%%%%%%%%%%%%%%%%%%%%%%%

\section{Interpretation and Modelling}\label{sec:inter}

The overall temporal and spectral evolution of the optical emission ($\Delta \alpha_{\rm opt} = 4.2$, $\Delta \beta_{\rm opt} = 3.3$; see Figure~\ref{fig:LC_GRB180618A}) are hard to explain just in terms of flaring activity (i.e. $\Delta t/t >1$; \citealt{2006ApJ...642..354Z}) or the external shock from the decelerating relativistic ejecta (i.e. the afterglow; \citealt{1999PhR...314..575P,2000ApJ...545..807K}). The observed decay rates of the emission do not agree with $\alpha \approx 1$ expected from forward shock emission with a typical electron index of the synchrotron energy spectrum, and the break cannot be reconciled with the passage of the synchrotron cooling frequency \citep{1998ApJ...497L..17S}. We also discard the fast-fading emission from the reverse shock \citep{2000ApJ...545..807K}, which typically decays with index $\alpha \approx 2$ ---still slower than in GRB 180618A. A reverse shock scenario further disagrees with the rapid spectral evolution of the optical emission at the time of the light curve break, and the non-detection of high values of polarization during the broad optical peak \citep{2013Natur.504..119M}. Finally, we rule out that the optical break is an effect of the relativistic collimation of the outflow given that there is no simultaneous steepening of the emission across the spectrum at that time \citep{1999ApJ...524L..43S,2009ApJ...698...43R}. 

We suggest that the spectral evolution and rapid decline of the optical emission of GRB 180618A is produced by thermal emission (e.g., \citealt{2011Natur.480...72T,2019Natur.565..324I}). In Section~\ref{sec:thermalemission}, we model the ultraviolet to optical emission with a black body model. Following, we discuss if the radioactive decay of the r-processed ejecta from the merger or a thermalized jet are credible interpretations to explain the origin of the bright thermal emission. In Section~\ref{sec:magnetar_nebula_relth}, we find evidence of a magnetar wind nebula powering thermal optical emission during $15-60 \,$minutes after GRB 180618A ---suggesting that emission before that time is due to the jet afterglow. In Section~\ref{sec:magnetar_nebula_nonth}, we detail the properties of the magnetar powering the distinct non-thermal emission components, from the gamma rays to optical wavelengths. In Section~\ref{sec:afterglow_nonth}, we model the overall (X-ray to optical) spectra with physical synchrotron models. This analysis proves Section~\ref{sec:magnetar_nebula_relth} findings; the emission during $3-15 \,$minutes post-burst is due to the jet afterglow and that corresponding to $15-60 \,$minutes post-burst requires an extra spectral component (i.e. thermal emission). In Section~\ref{sec:colli}, we test closure relations, and we suggest that the jet geometry and observer viewing angle are key to detecting the thermal emission. 

\subsection{Thermal Emission} \label{sec:thermalemission}

\begin{figure}[ht!]
\centering
\includegraphics[width=\columnwidth]{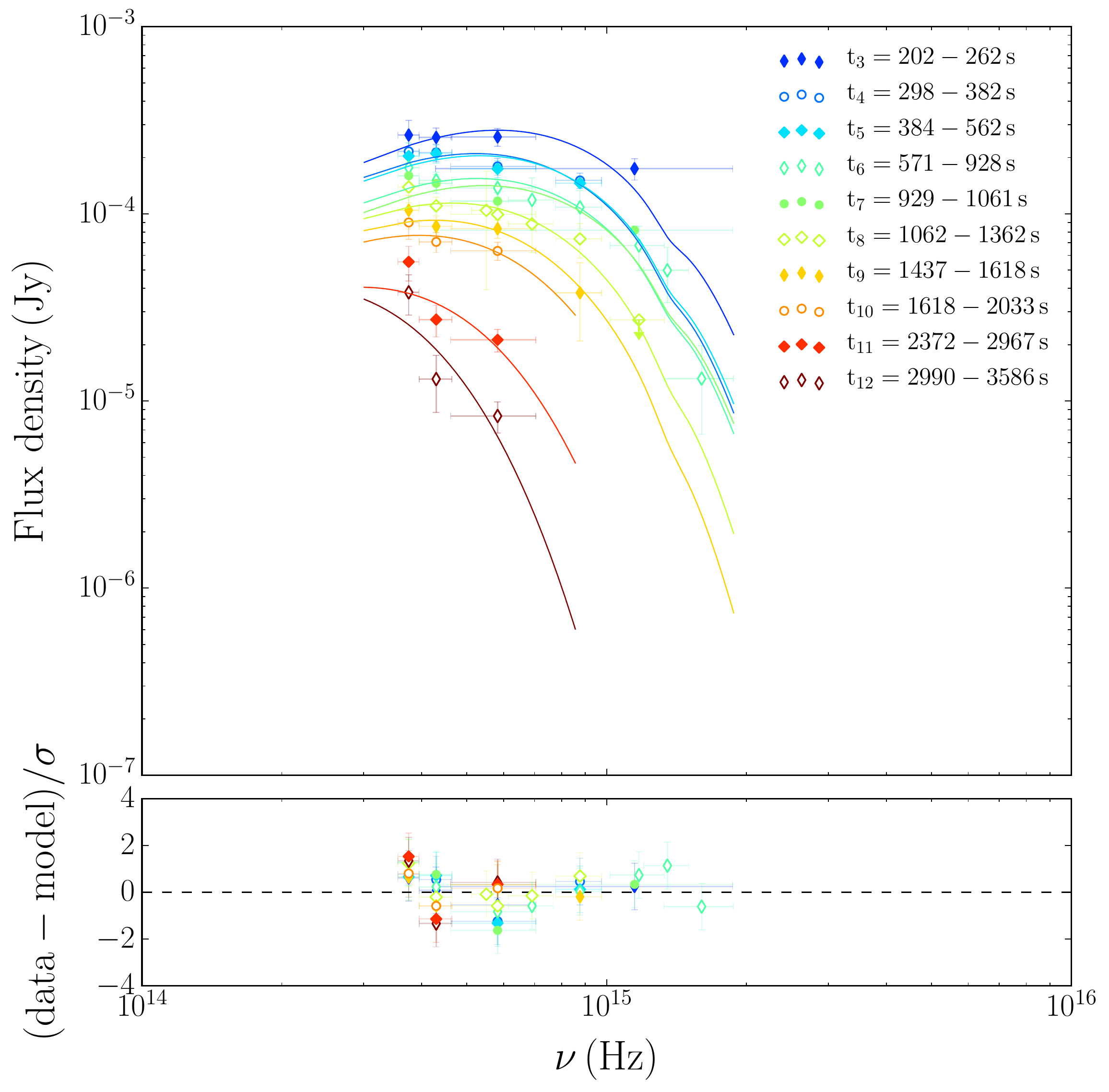}
\caption{The GRB 180618A best-fitting black body models to the joint RINGO3/UVOT data at the observer rest-frame. Note that the model accounts for Galactic dust extinction, but does not include the host galaxy dust contribution.}
\label{fig:thermal_emission1}
\end{figure}

In {\it Xspec}, we modelled the co-temporal UVOT and RINGO3 data with a black body profile that includes the Milky Way dust absorption (see Figure~\ref{fig:thermal_emission1}). At the observer rest-frame, the best-fitting effective temperatures and luminosities from the black body models evolve from $T_{\rm eff} = (1.1 \pm 0.1) \times 10^4 \,$K and $L_{\rm th}/D_{\rm L}^2 = (4.3 \pm 0.5) \times 10^{44} \,$erg s$^{-1}$ Gpc$^{-2}$ at $t=202-262\,$s post-burst to $T_{\rm eff} = (3.8 \pm 0.6) \times 10^3 \,$K and $L_{\rm th}/D_{\rm L}^2 = (1.8 \pm 0.6) \times 10^{43} \,$erg s$^{-1}$ Gpc$^{-2}$ at $t=2990-3586\,$s post-burst (see results in Table~\ref{tab:BB}), where $D_{\rm L}$ is the luminosity distance. Note that the data is well fitted by the model without the need of the host galaxy dust extinction component, which is consistent with the large offset of the short GRB 180618A with the host galaxy core (i.e. $\approx 10 \,$kpc).

Faint thermal emission at optical and infrared wavelengths has been detected emerging hours to days after short GRBs \citep{2013Natur.500..547T,2017ApJ...848L..12A} ---when the relatively brighter afterglow has subsided. This emission has been attributed to the kilonova \citep{2010MNRAS.406.2650M,2011ApJ...736L..21R} ---a type of emission powered by the radioactive decay of the heavy elements formed after the dynamically-ejected and wind-driven neutron-rich material of the merger expands from nuclear densities and neutrons are captured via the r-process. Predictions are that an ultraviolet kilonova precursor could also be detected from the decay of free neutrons in the fast material \citep{2015MNRAS.446.1115M}. However, the short timescales of the GRB 180618A optical emission do not support r-process models, for which we also expect lower peak luminosities ($L_{\rm th} = 10^{40}-10^{42} \,$erg$\,$s$^{-1}$; \citealt{2010MNRAS.406.2650M,2015MNRAS.446.1115M}) than those estimated in GRB 180618A, i.e. $L_{\rm th} (z=0.554) \approx 2 \times 10^{45} \,$erg$\,$s$^{-1}$.

We also discard thermal emission powered by the GRB jet. The measured luminosities and effective temperatures do not follow the scalings expected from thermal emission of the relativistic and non-relativistic material of the jet cocoon \citep{2017ApJ...834...28N,2018MNRAS.478.4553D}. Similarly, maximizing the signal from the jet cocoon at $z=0.554$ corresponds to a very energetic long GRB and large energy stored in the jet cocoon, on the order of $E_{\rm c} \approx 10^{52} \,$erg \citep{2017ApJ...834...28N}. This would be expected to produce at most a  $\approx 22 \,$mag near-ultraviolet and $\approx 23.6 \,$mag optical thermal rebrightening. These values are not consistent with the observed peak luminosities of the thermal emission (e.g., $\approx 18 \,$mag in the UVOT {\it uvw1} band, and $\approx 18.7 \,$mag in the RINGO3 {\it BV} band), and they are far below the limiting magnitude of our UVOT and RINGO3 observations. Furthermore, successful jets in energetic engines like the GRB 180618A are expected to have an early breakout from the stellar envelope, which cuts thermalization and leads to most of the jet energy leaving the ejecta \citep{2002MNRAS.337.1349R,2018ApJ...866....3D,2019Natur.565..324I}.

\subsection{Thermal Emission from a Relativistically-expanding Magnetar Wind Nebula} \label{sec:magnetar_nebula_relth}

Here, we explore the possibility that the observed thermal emission from the GRB 180618A could be explained by a magnetar wind nebula  \citep{2013ApJ...776L..40Y,2014MNRAS.439.3916M,2015ApJ...807..163G,2019ApJ...880...22W}, and the implications of such a model for this system.

In a magnetar wind nebula, the expanding ejecta shell is continually being heated from behind by the magnetar winds ---a scenario that differs from fireball models with energy injection at a single point in time (e.g. \citealt{1997ApJ...476..232M,2017ApJ...834...28N}). As such, the thermal luminosity is expected to track the spin-down luminosity of the magnetar as $L_{\rm th} \propto L_{\rm sd} \propto t^{-2}$ \citep{2013ApJ...776L..40Y,2014MNRAS.439.3916M}. Assuming that the ejecta radially expands as $R_{\rm ej} \propto t$, the effective temperatures are expected to follow $T_{\rm eff} = [L_{\rm th} / (4 \pi \sigma_{\rm B} R_{\rm ej} ^2)]^{1/4} \propto  L_{\rm th} ^{1/4} t^{-1/2} \propto t^{-1}$, where $\sigma_{\rm B}$ is the Stefan-Boltzmann constant.

\begin{figure}[ht!]
\centering
\includegraphics[width=\columnwidth]{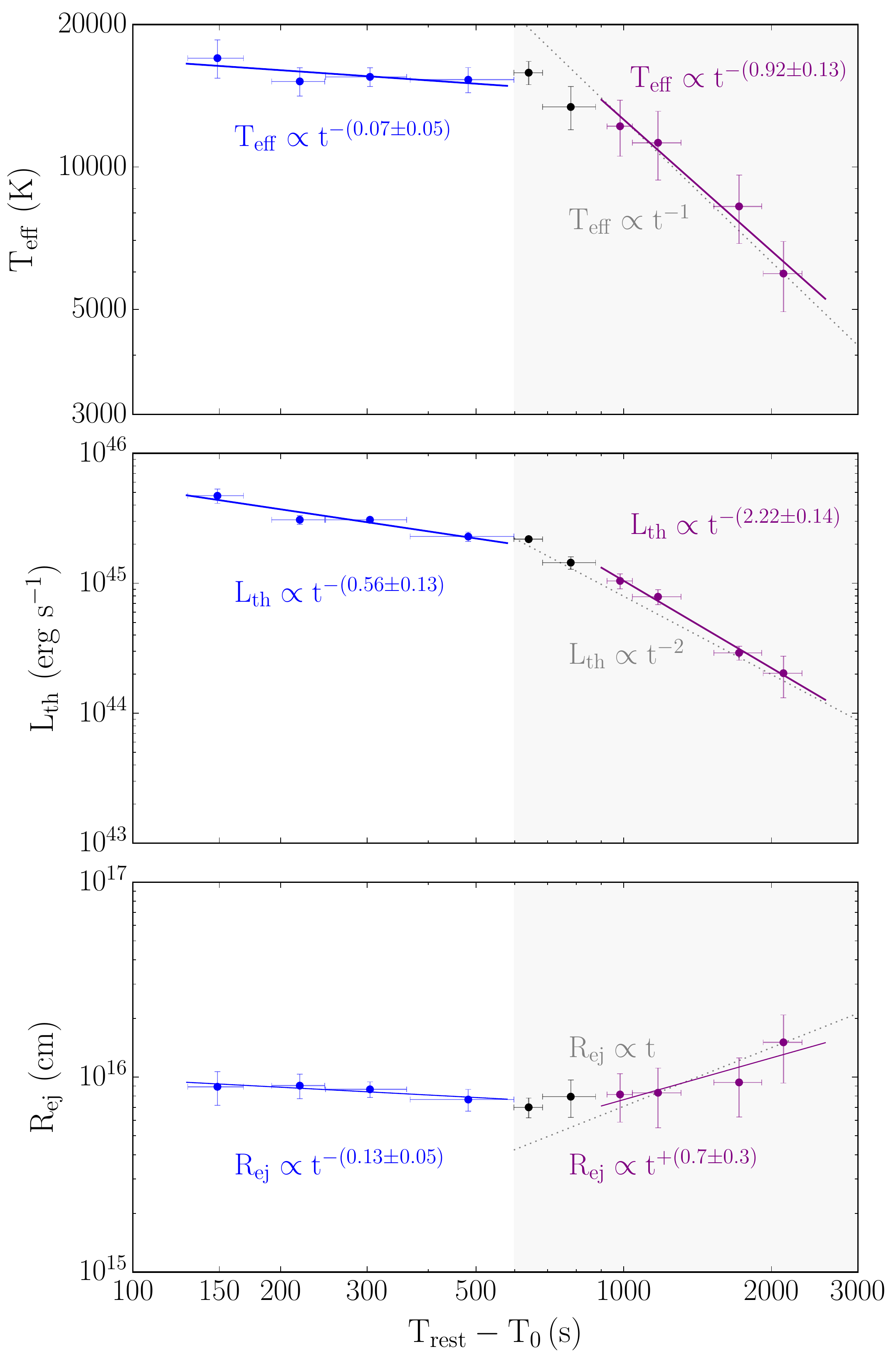}
\caption{Temporal evolution of the effective temperatures ($T_{\rm eff}$), luminosities ($L_{\rm th}$) and photospheric radii ($R_{\rm ej}$) of the best-fitting black body models at the host galaxy rest-frame (i.e. $z=0.554$). The shaded grey area corresponds to the thermal contribution from the magnetar wind nebula, and the dotted grey line is the expected evolution of the temperature, luminosity and radius for an optically thin regime.}
\label{fig:thermal_emission2}
\end{figure}

In Figure~\ref{fig:thermal_emission2}, we model the temporal evolution of the best-fitting effective temperatures, luminosities and photospheric radii ($R_{\rm ej}$) with power laws. For the data after $\approx 1400 \, $s post-burst, we find that the observed scalings $L_{\rm th} \propto t^{-(2.22\pm 0.14)}$, $T_{\rm eff} \propto t^{- (0.92 \pm 0.13)}$ and $R_{\rm ej} \propto t^{+(0.7 \pm 0.3)}$ are compatible with the expected in a magnetar nebula \citep{2014MNRAS.439.3916M}. The peak luminosity $L_{\rm th} (z=0.554) \approx 2 \times 10^{45} \, $erg$\,$s$^{-1}$ of the GRB 180618A is also within the values expected \citep{2013ApJ...776L..40Y,2014MNRAS.439.3916M}. Note that the data at $900-1400 \,$s post-burst are still in agreement with the model, and that the deviation we detect is likely due to the contribution from the external shock, which dominates the emission before $900 \,$s post-burst (see Section~\ref{sec:afterglow_nonth}).

A new-born millisecond magnetar from a binary neutron star merger is expected to be close to the centrifugal breakup limit \citep{2013ApJ...771L..26G,2015ApJ...812...24F}, with an initial spin period of $P_{\rm i} \approx 1\,$ms. Consequently, ejecta below a critical mass $M_{\rm ej } \approx 10^{-2} M_{\odot} P_{\rm i}^{-2}$ can be accelerated to trans-relativistic speeds ($\Gamma \geq 1$) by the magnetar wind \citep{2013ApJ...776L..40Y,2013ApJ...771...86G}, where $\Gamma=(1-\beta^2)^{-1/2}$ is the Lorentz factor, and $\beta=v/c$ is the ratio of the velocity between inertial reference frames and the speed of light in vacuum. From the photospheric radius of the best-fitting black body model at the engine rest-frame, $R_{\rm ej}^{\prime} = c t' \approx  c t_{\rm obs} (1+z)^{-1} (1-\beta)^{-1} \approx 2 \Gamma^2 c t_{\rm obs} (1+z)^{-1}$, we find that the ejecta in the GRB 180618A system is expanding at mildly relativistic speeds, i.e. $\Gamma (z=0.554) \approx 9$.

For such low ejecta masses giving rise to high Lorentz factors, the ejecta can become optically thin before the diffusion time \citep{2013ApJ...776L..40Y}, i.e. when the optical depth $\tau^{\prime} = (3 M_{\rm ej }  \kappa) / (4 \pi R^{\prime \, 2}) \approx 1$, where $\kappa$ is the opacity and $M_{\rm ej }$ is the isotropic-equivalent mass of the ejecta. The optical depth can be significantly raised (for a total mass in the shell) given electron-positron pair creation in the region behind the ejecta shell \citep{2014MNRAS.439.3916M}. We estimate the maximum pair multiplicity by assuming that a fraction $\approx 0.1$ of the magnetar rotational energy $E_{\rm rot} \approx 10^{52}\,$erg is converted through a pair cascade process into electron-positron pairs in the nebula \citep{2014MNRAS.439.3916M}, such that $\kappa_{\pm} \approx (0.1 E_{\rm rot}/m_{\rm e} c^{2}) (m_{\rm p}/M_{\rm ej}) \approx 10^{3}(M_{\rm ej}/10^{-3}M_{\odot})^{-1}$. That is, pairs contribute $\approx 10^3$ times to the opacity for their mass than an electron-ion outflow, with typical electron-scattering opacity $\kappa \approx 0.2 \,$cm$^{2}\,$g$^{-1}$ \citep{2013ApJ...776L..40Y,2015MNRAS.446.1115M}.  Given an electron-positron annihilation rate slower than the outflow expansion speed at late times \citep{2014MNRAS.439.3916M}, the pair opacity will still dominate the total opacity by the time of observations. Therefore, assuming an optically thin regime ($\tau^{\prime} \lesssim 1$) when we start to notice the thermal emission ($t_{\rm obs} \lesssim 10^3 \,$s) and opacity $\kappa  \approx (0.2 \,$cm$^{2}\,$g$^{-1} ) \, \kappa_{\pm}$, we estimate ejecta masses $M_{\rm ej } \lesssim 10^{-4} M_{\odot} (\kappa_{\pm}/10^3)^{-1} $ at the polar regions of the system. This is consistent with the mass-loss expected from merger remnants \citep[e.g.][]{2006MNRAS.368.1489O,2007NJPh....9...17L}.

\subsection{Non-thermal Emission from a Magnetar Wind Nebula} \label{sec:magnetar_nebula_nonth}

\begin{figure*}[ht!]
\centering
\plottwo{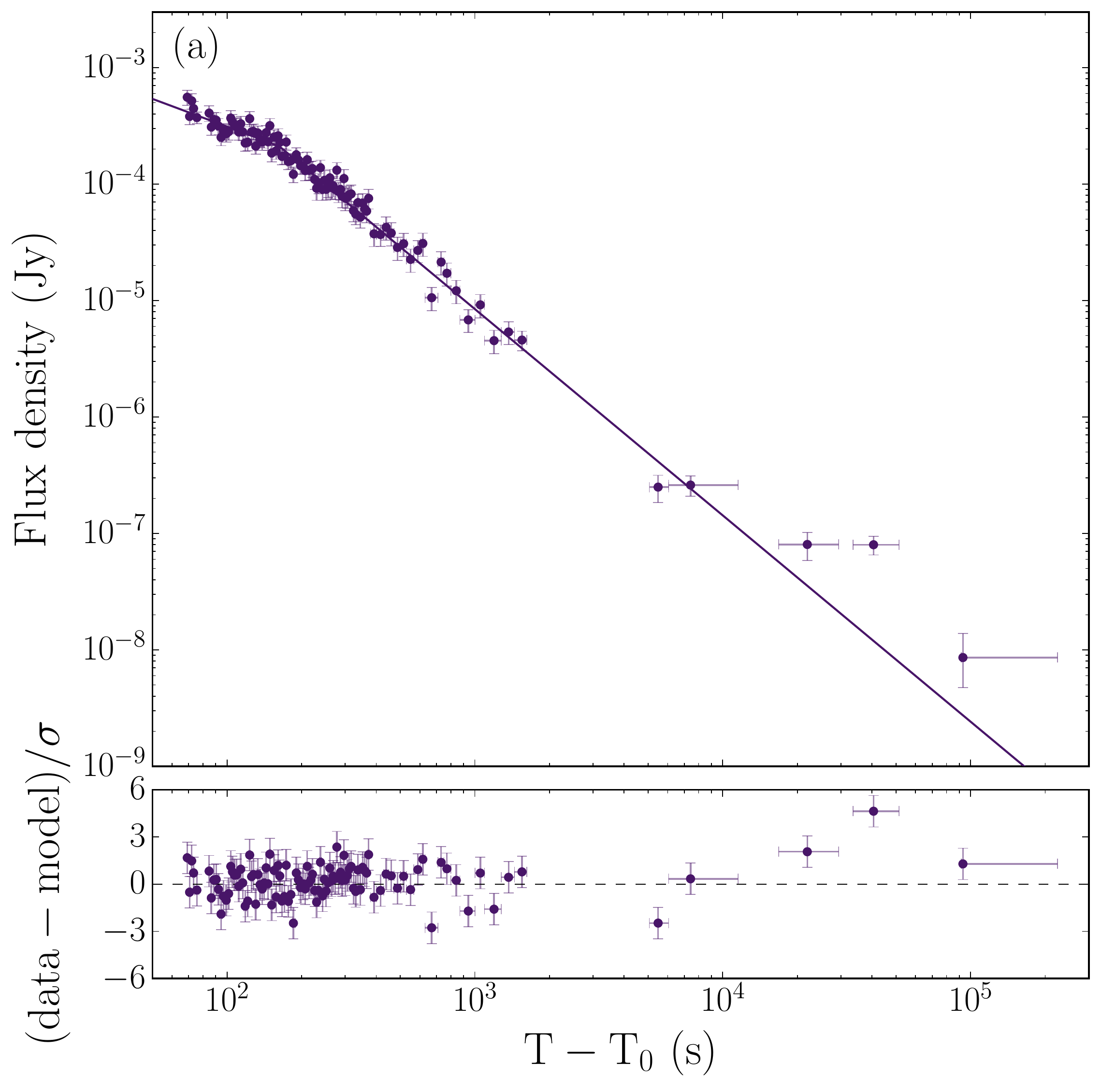}{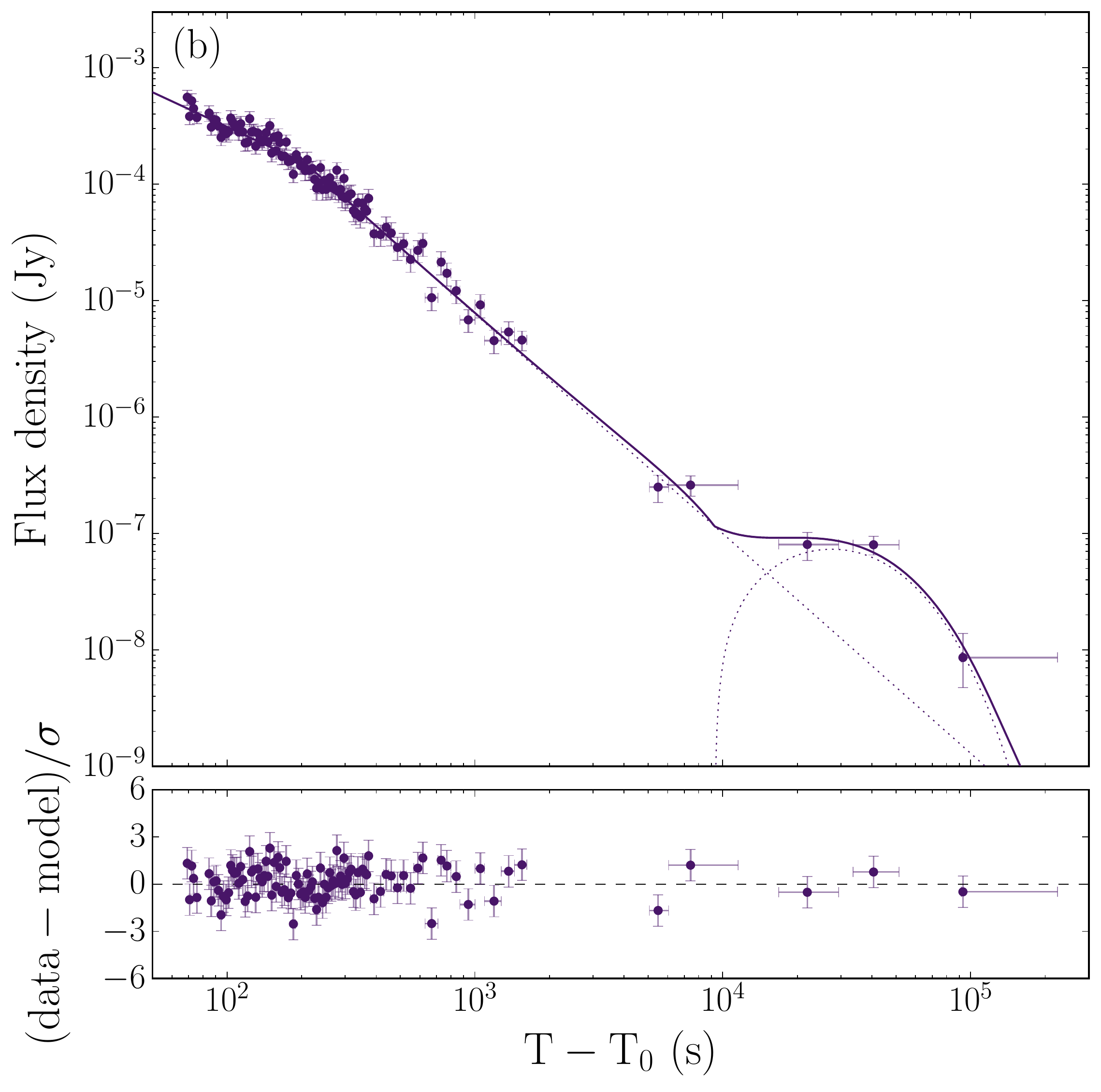}
\caption{Best-fitting models of the XRT X-ray light curve of the GRB 180618A. (a) Broken power-law model. The best-fitting parameters are an initial decay rate $\alpha_{\rm X, 1} =  0.79 \pm 0.12$, steeping to $\alpha_{\rm X, 2} = 1.77 \pm 0.04 $ at $ t_{\rm break} = 162 \pm 11 \,$s ($\chi^2/$dof$ = 131/102$). (b) Broken power-law model plus a pulse function, with $\alpha_{\rm X, 1} =  0.99 \pm 0.07$, $\alpha_{\rm X, 2} = 1.89 \pm 0.06 $, $ t_{\rm break} = 206 \pm 14 \,$s ($\chi^2/$dof$ = 104/99$). The bottom panels are the residuals of the best-fitting model.}
\label{fig:LC_180618A_Xrays}
\end{figure*}

We used the GRB 180618A pipeline-processed products of the Swift X-Ray Telescope (XRT; \citealt{2005SSRv..120..165B,2009MNRAS.397.1177E}), and we modelled the 0.3$-$10 keV light curve with a broken power-law function (see Figure~\ref{fig:LC_180618A_Xrays}-a). We find a significant emission excess of 4.6$\sigma$ at $\approx 4\times 10^4\,$s post-burst, which we modelled with a pulse function \citep{2008A&A...491..183P}, i.e. $F = F_0 (t - t_0) e^{-(t - t_0) / \tau_{\sigma}}$. The best-fitting parameters of this model are an initial emission decay rate  $\alpha_{\rm X, 1} =  0.99 \pm 0.07$, and a break at $ t_{\rm break} = 206 \pm 14 \,$s post-burst followed by a steeper decay with $\alpha_{\rm X, 2} = 1.89 \pm 0.06 $ (see Figure~\ref{fig:LC_180618A_Xrays}-b). For the late-time rebrightening, we find $t_0 = (9.3 \pm 3.8)\times 10^3 \, $s and $\tau_{\sigma}= (1.8 \pm 0.3) \times 10^{4} \,$s.

If the magnetar winds are powerful enough and the ejecta mass is low, the nebula can be fully ionized at late times and X-ray emission will be able to leak out the nebula \citep{2014MNRAS.439.3916M,2015ApJ...807..163G}. This is consistent with the X-ray late-time rebrightening of the GRB 180618A, and implies that we are directly detecting the magnetar spin-down luminosity. This emission is expected to follow $L_{\rm sd} = L_{0} (1+ t/t_{\rm sd})^{-2}$ with typical values $L_0 = 1.7 \times 10^{50} \, B_{15} ^2  P_{\rm i, -3} ^{-4} \,\,$erg$\,$s$^{-1}$ and $t_{\rm sd} = 307 \, B_{15} ^{-2}  P_{\rm i, -3}^{2}\,\,$s (e.g., \citealt{2019LRR....23....1M}), which correspond to a cooled neutron star of $12\,$km fiducial radius \citep{2020ApJ...888...97B}, magnetic field $B_{15} \approx 10^{15} \,$G and initial spin $P_{\rm i, -3} \approx 1 \,$ms \citep[see e.g.][]{2003MNRAS.345.1077R}. 

At early times ($t \ll t_{\rm sd}$), the luminosity of the magnetar from the GRB 180618A is estimated to be $L_0\approx L_{\rm iso} f_{\rm b}/\eta$, where $f_{\rm b}\approx \theta_{\rm j, EE}^2 /2$ is the beaming factor, $\eta$ is the radiative efficiency, and $\theta_{\rm j, EE}$ is the jet opening angle. For a redshift $z=0.554$ and assuming a large opening angle for the magnetar wind  (i.e. $f_{\rm b}/\eta \approx 1$; \citealt{2012MNRAS.419.1537B}), the characteristic luminosity is $L_0 (z=0.554) \approx 2 \times 10^{49} \,$erg$\,$s$^{-1}$. Furthermore, for a rebrightening of $L_{\rm X} (z=0.554) \approx 4 \times 10^{44}  \,$erg$\,$s$^{-1}$ at $t \approx 5 \times 10^{4} \, (1+z)^{-1} \,$s post-burst, the characteristic spin-down time of the magnetar needs to be $t_{\rm sd} \approx 200 \,(1+z)^{-1} \,$s. Note that this is consistent with the early optical plateau of the UVOT {\it white} band that we detect with decay rate $\alpha_{\rm opt, white}=0.12 \pm 0.08$ at $ \lesssim 200 \,$s post-burst (see also \citealt{2013MNRAS.430.1061R,2017A&A...607A..84K}). Given all these constraints, we estimate the initial spin of the magnetar remnant of the GRB 180618A to be $P_{\rm i}\approx 4 \,$ms, with magnetic field $B\approx 6\times 10^{15} \,$G.

In the magnetar scenario, a multiwavelength rebrightening could also be detectable hours to days post-burst given the deceleration of the mildly relativistic ejecta by the circumburst medium \citep{2013ApJ...771...86G}. At redshift $z=0.554$, testing this prediction would have required sensitive late-time observations (e.g. \citealt{2009ApJ...696.1871P}). That is, given typical values of circumburst medium for a $\approx10 \,$kpc GRB-host galaxy offset ($n \lesssim 10^{-3}\, $cm$^{-3}$; \citealt{2015ApJ...815..102F}), the best-case scenario for the GRB 180618A corresponds to an emission excess with peak luminosity $\lesssim 10^{-9} \,$Jy at X-ray bands, $\lesssim 25 \,$mag at optical bands, and $\lesssim 100\, \mu$Jy at radio bands \citep{2013ApJ...771...86G}.

\begin{figure*}[ht!]
\centering
\plottwo{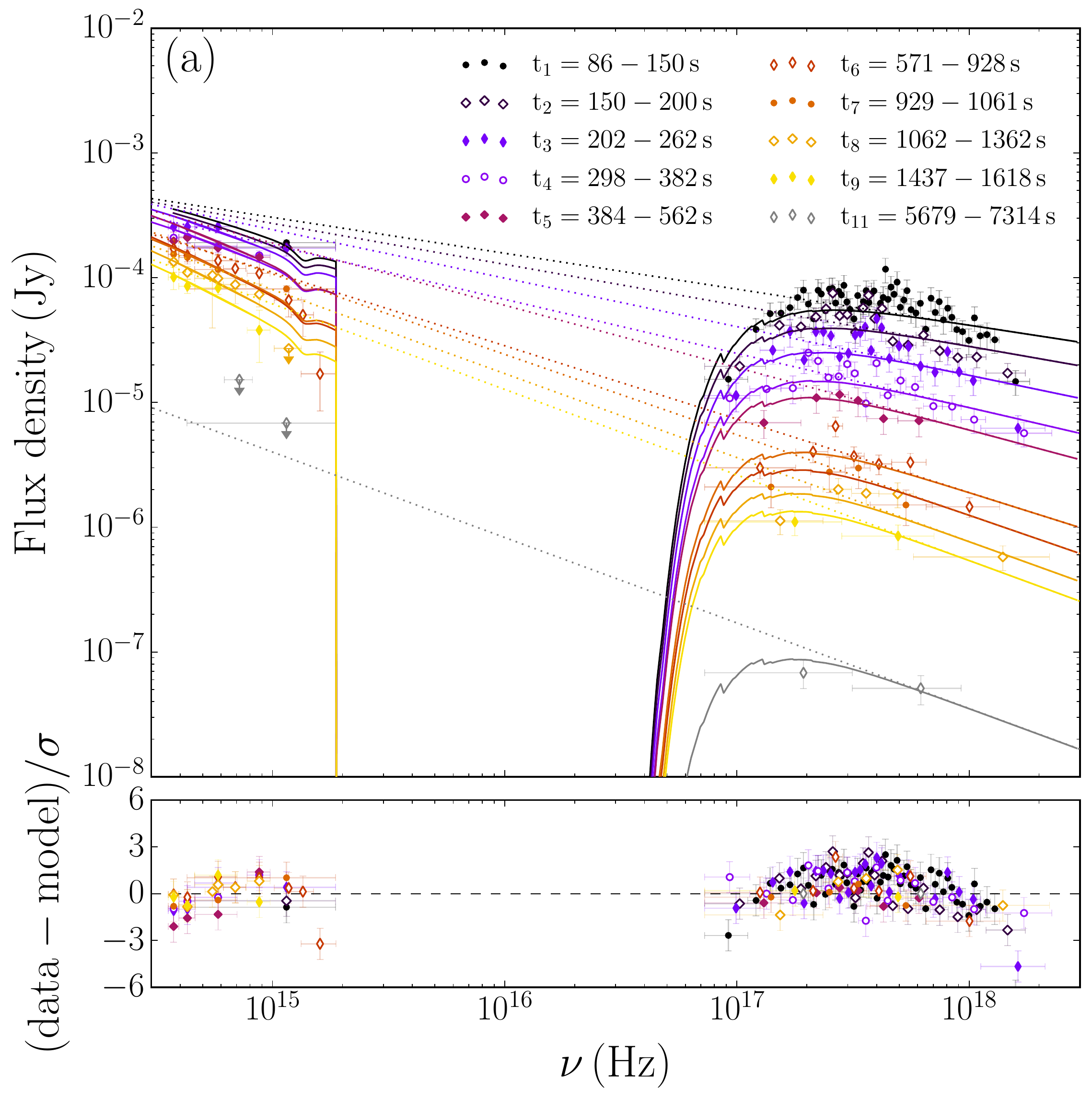}{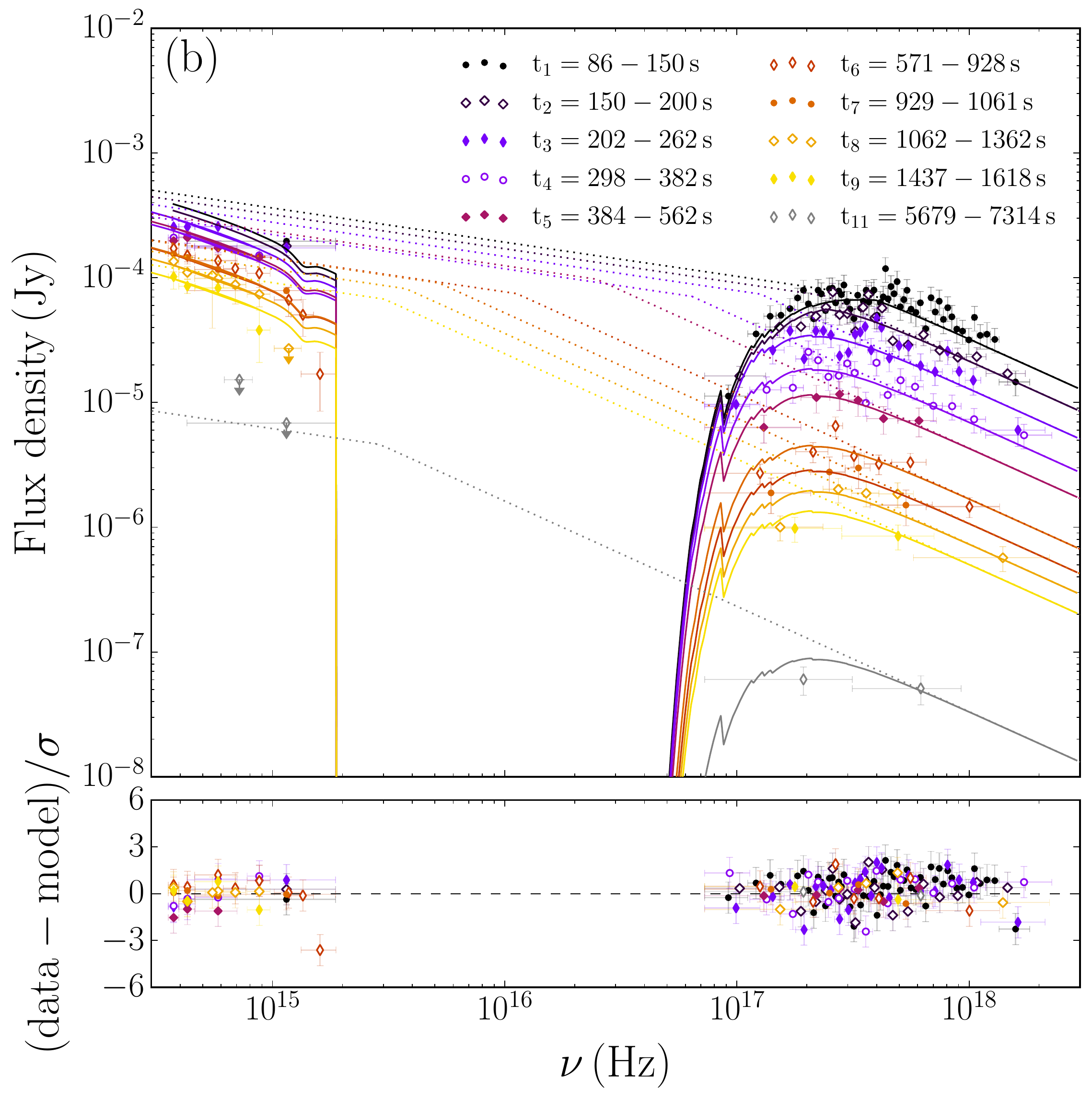}
\caption{Spectral energy distributions of the cotemporal X-ray, optical and ultraviolet observations of the GRB 180618A. The XRT, UVOT and RINGO3 data are modelled with the following synchrotron models. (a) Power laws ($\chi^2/$dof$ = 292/159$). (b) Broken power laws with a joint optical and X-ray spectral indexes ($\chi^2/$dof$ = 170/158$). The best-fitting spectral indexes are $\beta_{\rm opt}=0.27 \pm 0.02$ and $\beta_{\rm X}=0.84 \pm 0.05$. Note that we have fixed the break of $t_{11}$ epoch (grey color) to the $t_{10}$ epoch value given the optical upper limits. 
In dotted lines, we show the best-fitting synchrotron model and, in solid lines, the model that includes the dust and hydrogen absorption from the Milky Way and host galaxy for a redshift $z=0.554$. Detections have $1\sigma$ error bars, and non-detections are presented as $3\sigma$ upper limits.}
\label{fig:SEDs_180618A_fits1}
\end{figure*}

\begin{figure*}[ht!]
\centering
\plottwo{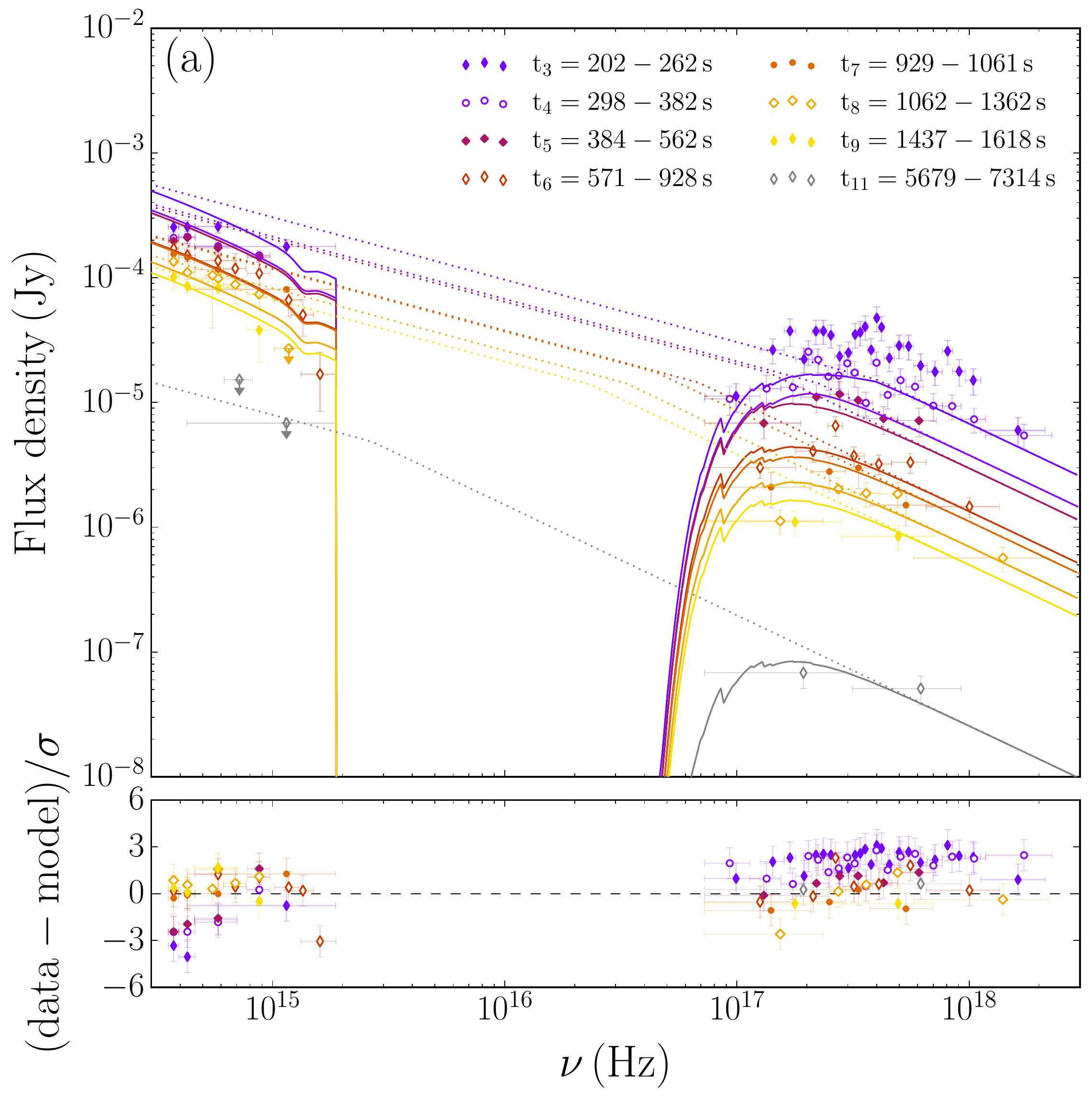}{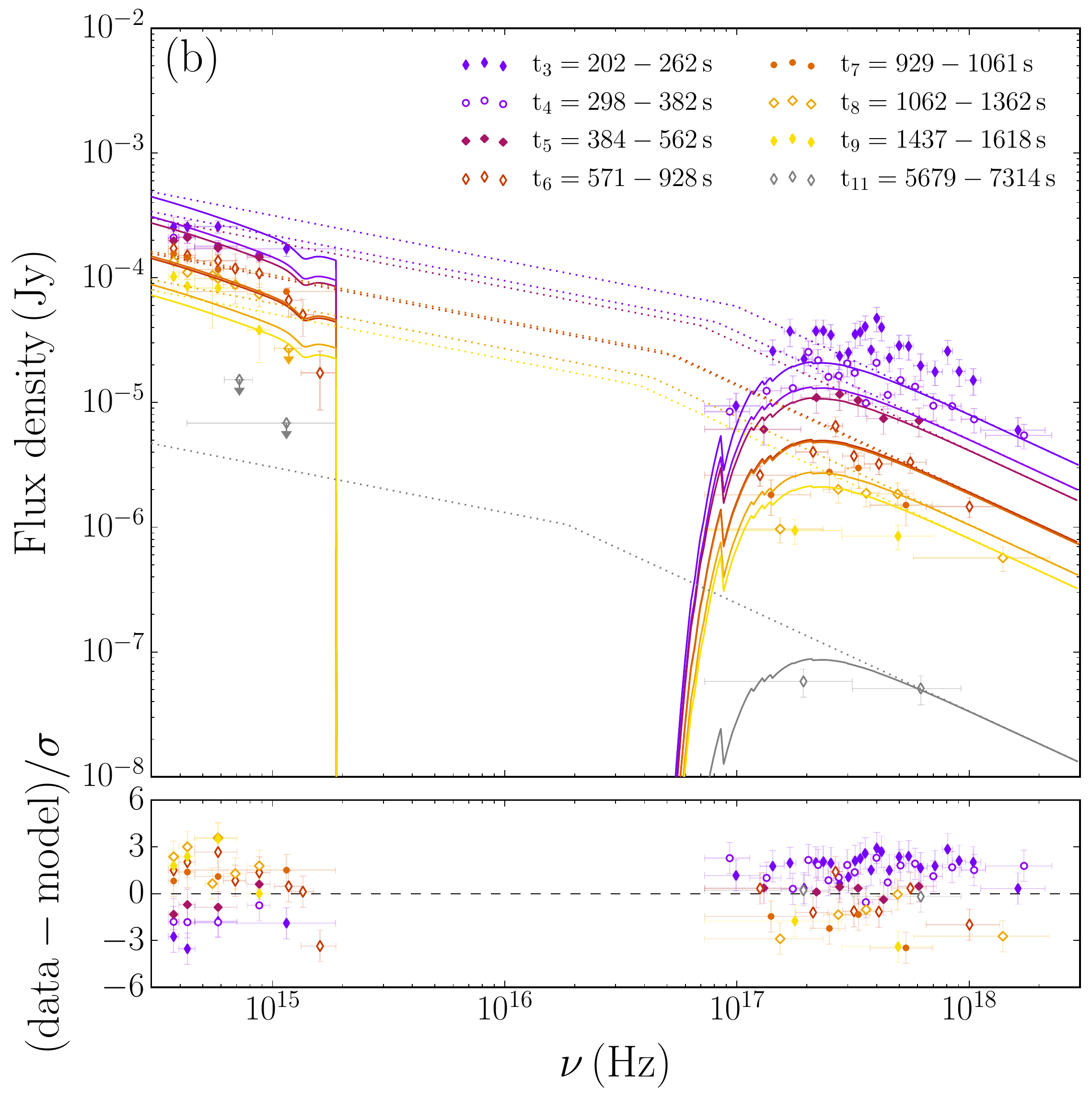}
\plottwo{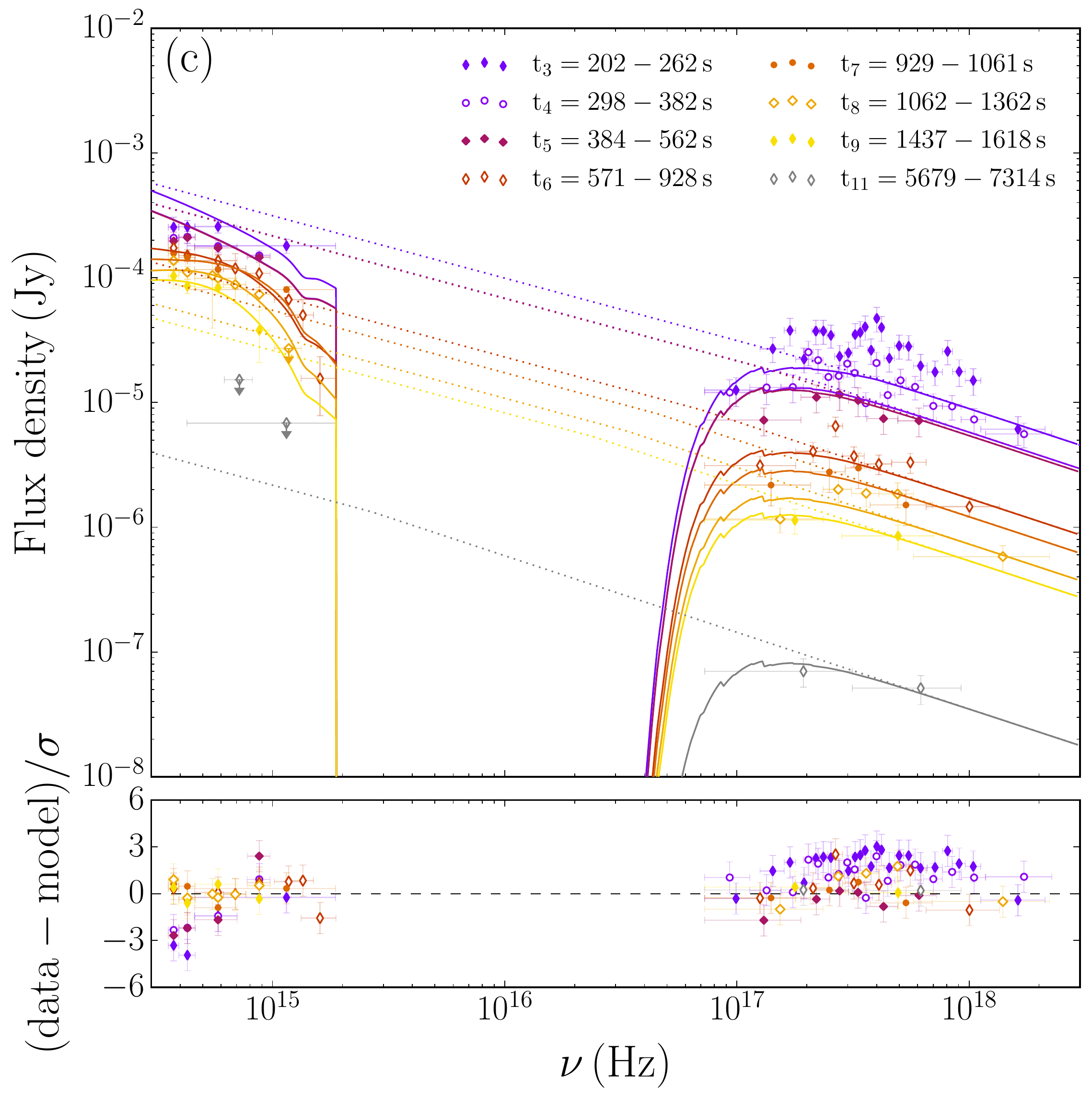}{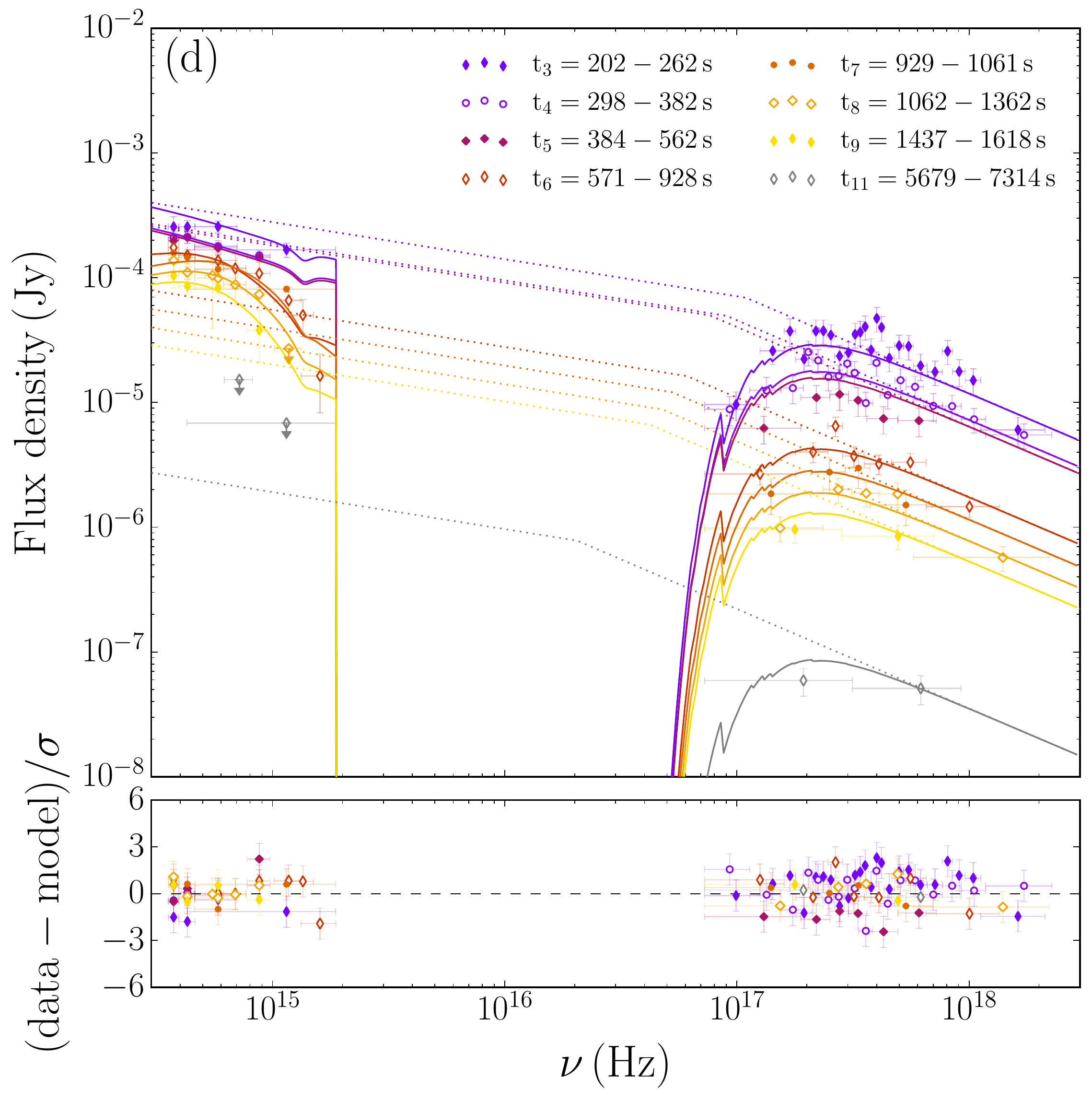}
\caption{Spectral energy distributions of the cotemporal X-ray, optical and ultraviolet observations of the GRB 180618A. The data after the {\it white}-band UVOT plateau (i.e., $\gtrsim 200 \,$s post-burst) are modelled with the following physical models. (a) Fast cooling of the electrons ($\chi^2/$dof$ = 309/90$). That is, we fix the optical spectral index to $\beta_{\rm opt}=\beta_{\rm opt, PI}-1=0.5$ and we let the synchrotron frequency evolve as $\nu_{\rm m} \propto t^{-1.5}$. (b) Slow cooling of the electrons ($\chi^2/$dof$ = 330/90$), i.e. we fix $\beta_{\rm X}=\beta_{\rm opt} + 0.5$ and the cooling frequency $\nu_{\rm c} \propto t^{-0.5}$. (c) Fast cooling and a black body profile ($\chi^2/$dof$ = 231/82$).  (d) Slow cooling and a black body profile ($\chi^2/$dof$ = 106/82$). The best-fitting optical spectral index is $\beta_{\rm opt}=0.30 \pm 0.08$ ($\beta_{\rm X}=\beta_{\rm opt} + 0.5$), which corresponds to an electron index $p=1.6 \pm 0.1$. The dust contribution from the host galaxy is estimated to be E($B-V$)$_{\rm HG}<0.02$.}
\label{fig:SEDs_180618A_fits2}
\end{figure*}

\subsection{Non-thermal Emission from the Afterglow} \label{sec:afterglow_nonth}

The deceleration of the relativistic ejecta by the circumburst medium should also have left an early imprint on the overall emission. To reproduce the synchrotron spectrum of this external shock \citep{1999PhR...314..575P}, we modelled the joint optical and X-ray spectral energy distributions with absorbed power laws and broken power laws in {\it Xspec} (see Figure~\ref{fig:SEDs_180618A_fits1}-a,b). Given the spectral slopes and the progression of the break to lower frequencies, we find that there must be at least one break frequency in between the bands and that, it must be either the synchrotron or cooling frequency in an interstellar medium \citep{2000ApJ...536..195C}. 

For the data after the X-ray light curve break and the optical plateau, we tried two physical models for the GRB synchrotron spectrum: the fast and slow cooling of the electrons (e.g., \citealt{1999PhR...314..575P}). The best-fitting synchrotron models display trends in the residuals, and suggest an additional spectral component at optical bands (see Figure~\ref{fig:SEDs_180618A_fits2}-a,b). Therefore, having in mind Section~\ref{sec:magnetar_nebula_relth} findings, we introduced a black body profile in the model (see Figure~\ref{fig:SEDs_180618A_fits2}-c,d). The best-fitting synchrotron plus black body model suggests a synchrotron spectrum with a rather hard electron index ($p= 1.6\pm 0.1$ with spectral slopes $\beta_{\rm opt}= \beta_{\rm opt, PI} -1= 0.30 \pm  0.06 $ and $\beta_{\rm X}= \beta_{\rm opt} + 0.5$), an interstellar medium profile, slow cooling regime, the cooling frequency in between the optical and X-ray bands, a black body profile contributing from $\approx900 \,$s post-burst, and low host galaxy dust extinction, i.e. E($B-V$)$_{\rm HG}<0.02$ ---consistent with the GRB 180618A-host galaxy large offset. Additionally, we note that the constraint on polarization of $P_{\rm BV} <6.1\%$ at early times supports the scenario of unpolarized forward shocks in short GRBs (see e.g. \cite{2021MNRAS.505.2662J} for long GRBs).

This physical model suggests that when the thermal emission subsides, we should detect the underlying afterglow again. At $\approx 3300 \,$s post-burst, the fast-fading afterglow emission should be $F_{\nu} \approx 7 \times 10^{-6} \,$Jy, which we speculate that it is consistent with the subtle light curve flattening of the RINGO3 {\it BV} band, with $F_{\nu}= (9 \pm 3) \times 10^{-6} \,$Jy. Furthermore, the emission across optical and X-ray bands is decaying faster than the expected for a spherical expansion \citep{2009ApJ...698...43R}, and suggests a collimated relativistic outflow.

\subsection{A Collimated Outflow} \label{sec:colli}

The temporal and spectral properties of the early X-ray emission of the GRB 180618A satisfy closure relations for a jetted outflow with a hard electron index. That is, the first light curve segment corresponds to the normal spherical decay rate of the afterglow (with average $\alpha_{\rm X} \approx 1.2$), and the second segment to the post-jet-break decay ($\alpha_{\rm X} \approx 2$; \citealt{2009ApJ...698...43R}). For an interstellar medium profile, slow cooling regime and with the X-ray band above the cooling frequency, a mean spectral index $\beta_{\rm X} = 0.80 \pm 0.06$ implies an electron index $p=1.6 \pm 0.1$; see also the temporal evolution of $\beta_{\rm X}$ in Figure~\ref{fig:LC_GRB180618A}-bottom panel. We find a temporal slope $\alpha_{\rm X, 1}= (3\beta_{\rm X} + 5)/ 8  = 0.93 \pm 0.02$ for the normal spherical decay of the afterglow, and $\alpha_{\rm X, 2} = (\beta_{\rm X} + 3)/ 2 = 1.90 \pm 0.03 $ after the light curve jet break, for an uniform jet scenario with lateral spreading \citep{2004IJMPA..19.2385Z,2009ApJ...698...43R} ---consistent with the observed decay rates of the X-ray light curve. Taking into account the early jet break at $t \approx 200 \,$s post-burst \citep{1999ApJ...519L..17S} and that we are detecting the bright gamma-ray prompt emission \citep{2002ApJ...571L..31Y}, we suggest that the observer faces the jet with a line of sight that likely runs near the jet edge. Overall, the GRB 180618A suggests that the degree of collimation \citep{2015ApJ...815..102F} and the observer viewing angle are key in detecting the short-lived optical thermal emission (and the late-time X-ray rebrightening) in short GRBs.

\begin{figure*}[ht!]
\centering
\includegraphics[width=35mm]{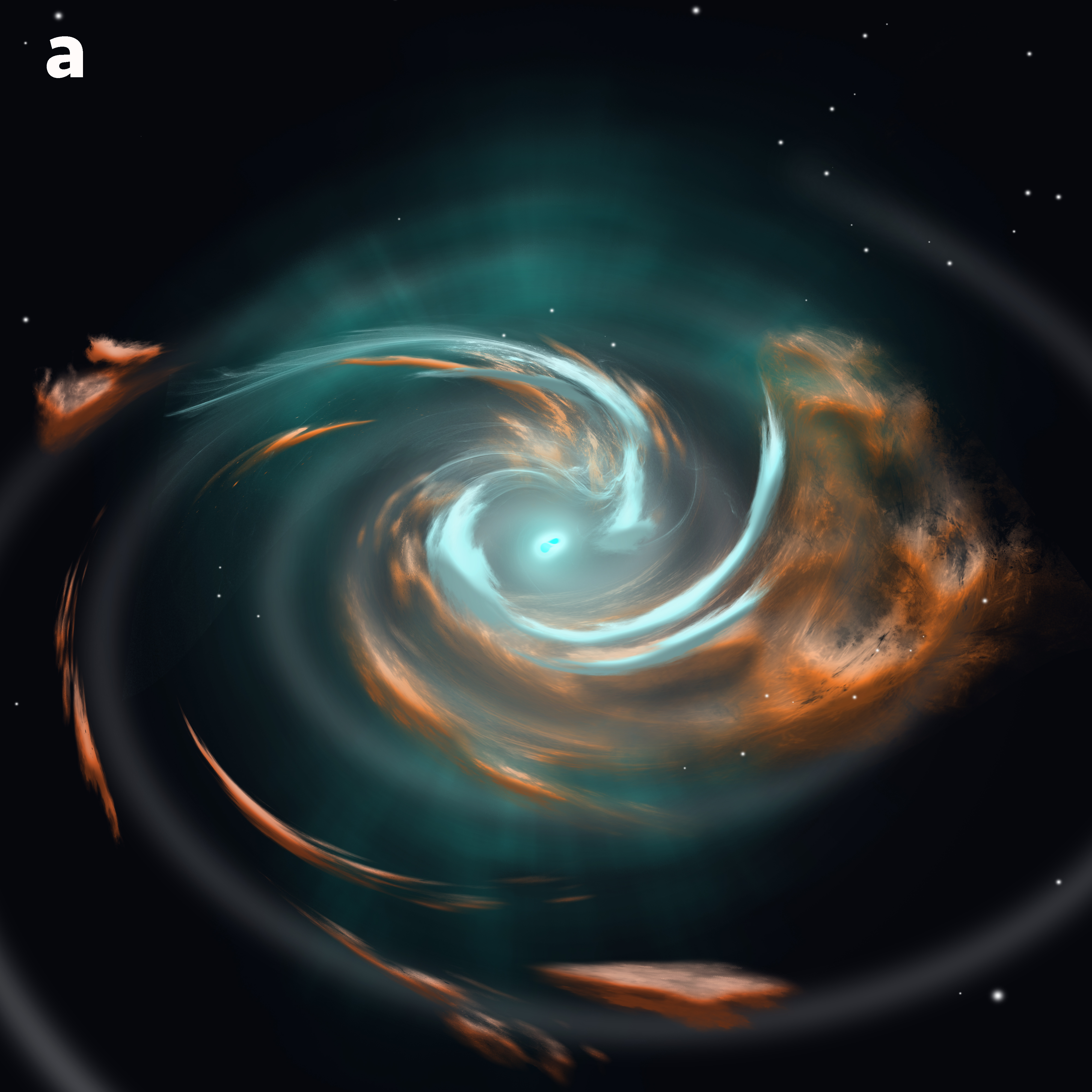}
\includegraphics[width=35mm]{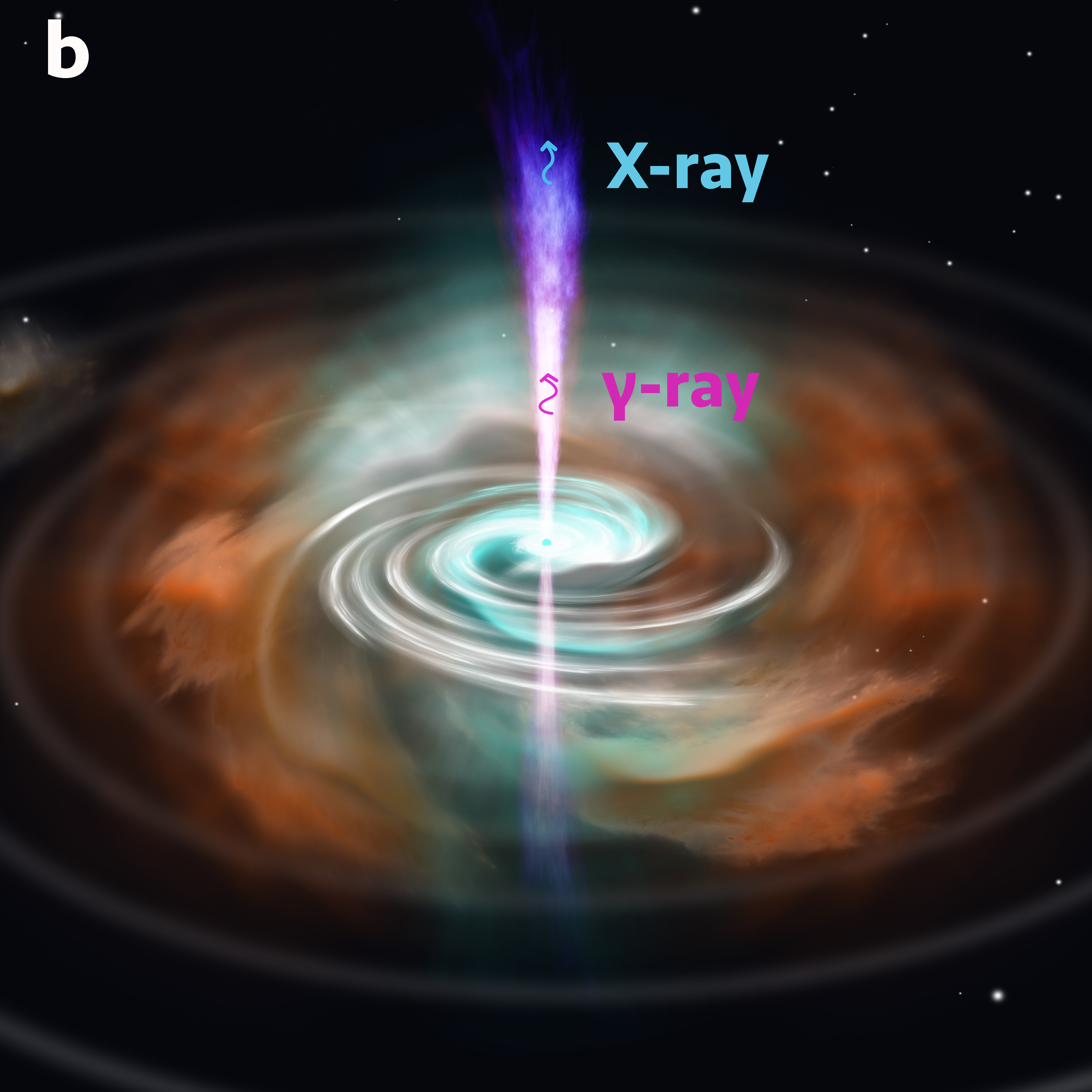}
\includegraphics[width=35mm]{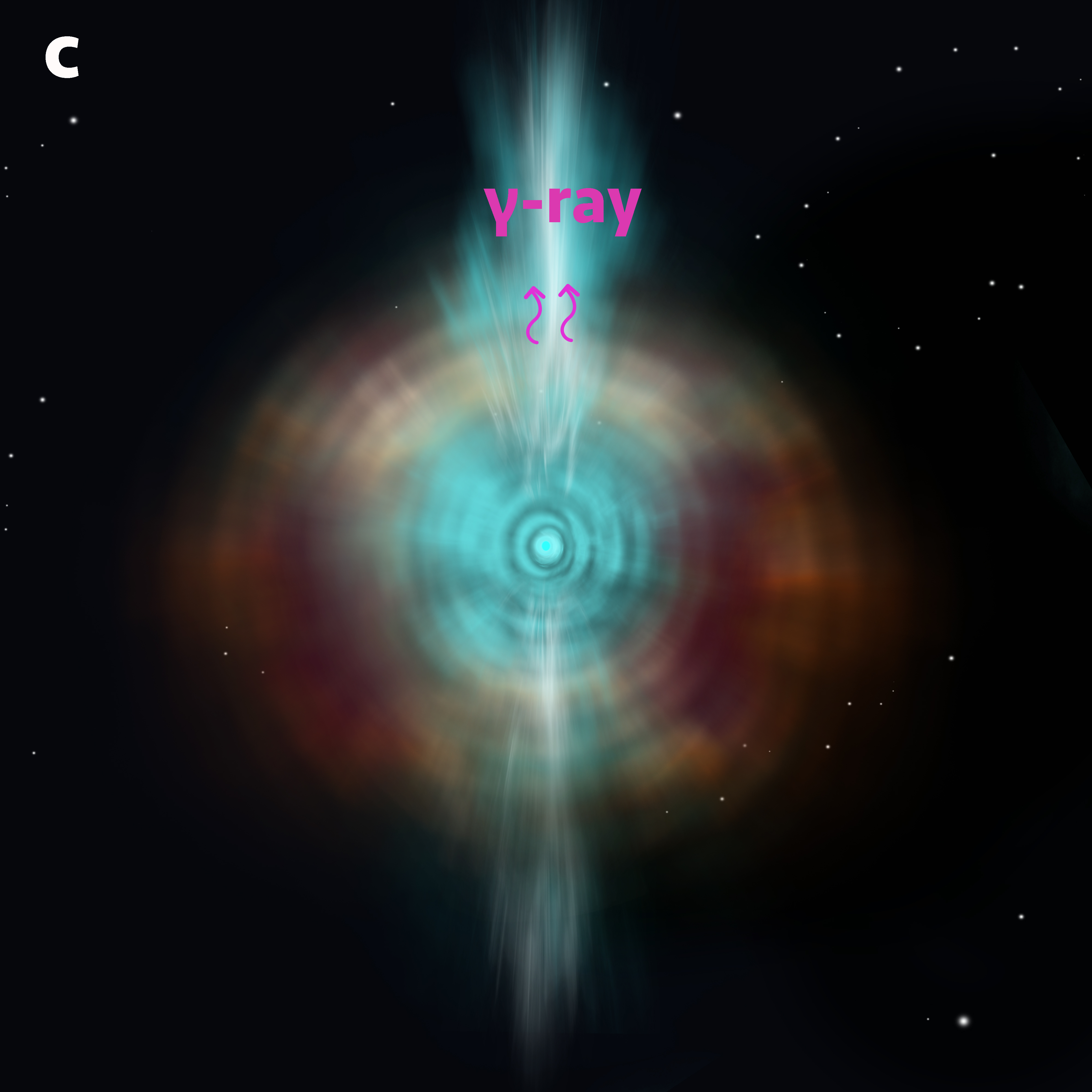}
\includegraphics[width=35mm]{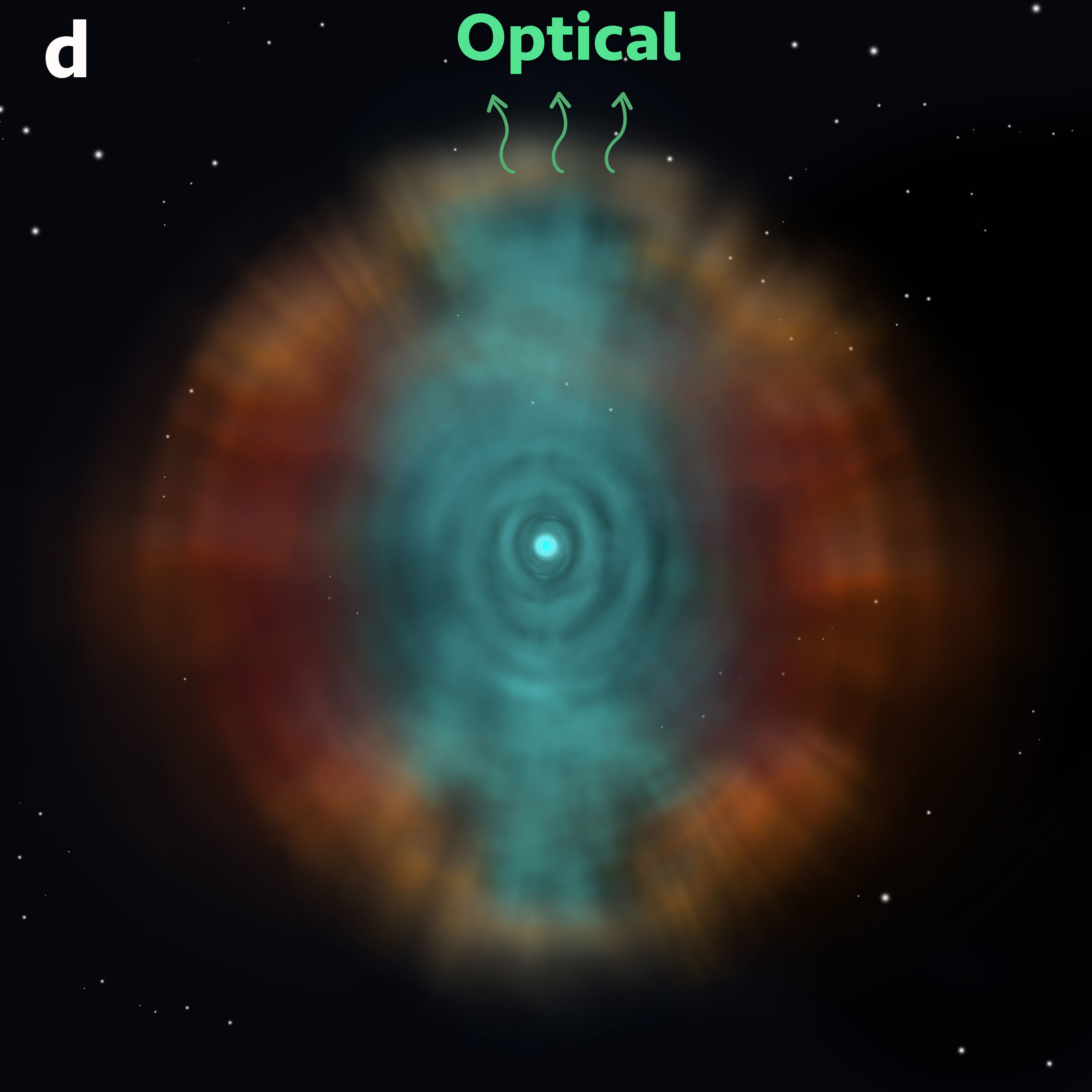}
\includegraphics[width=35mm]{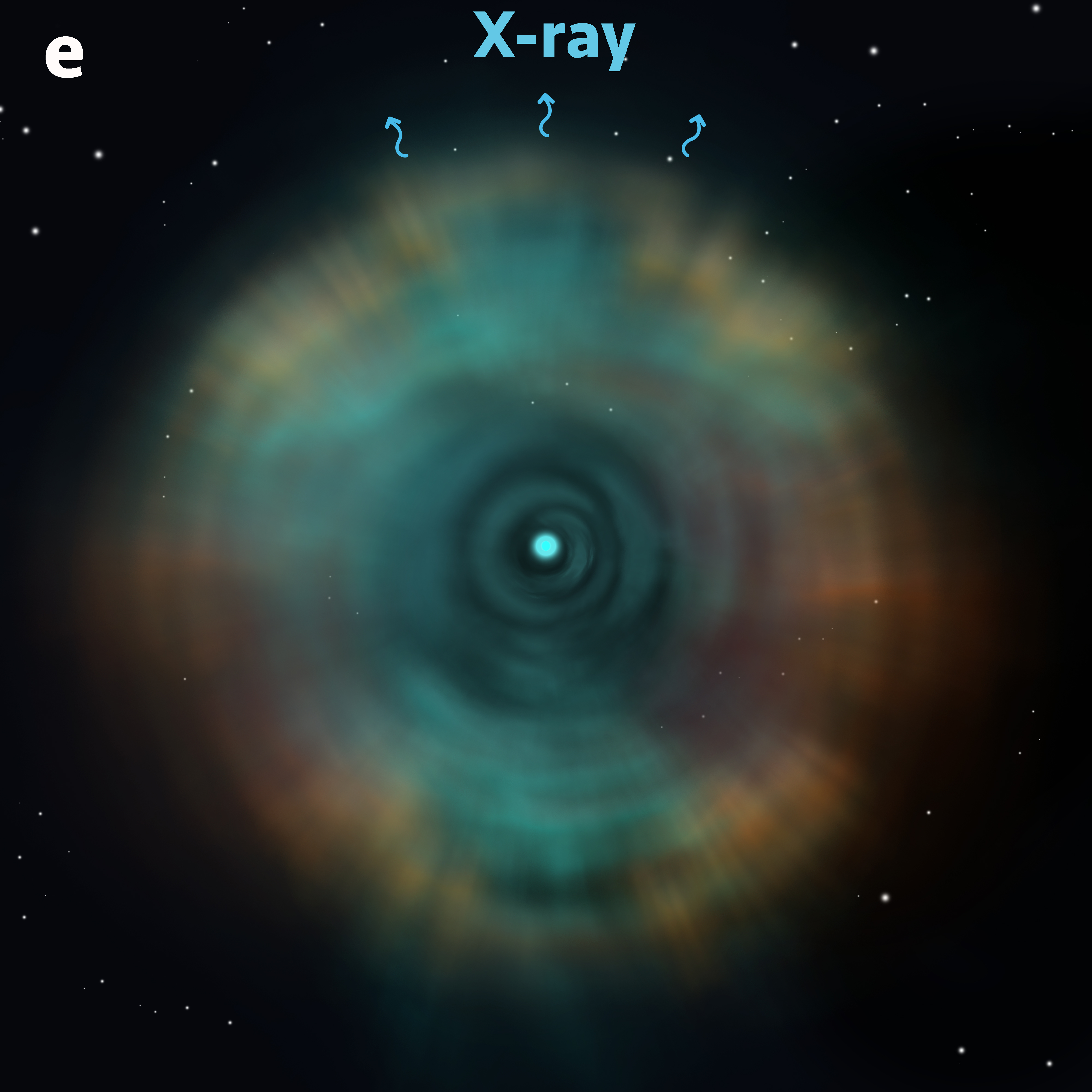}
\caption{Artist impression of the different energy sources powering the GRB 180618A multiwavelength emission. (a) The material is equatorially ejected by tidal forces during the neutron star binary merger \citep{2011ApJ...736....7C}, and radially ejected by hydrodynamic interactions at the neutron stars contact region (e.g., \citealt{2019LRR....23....1M}). (b) The accretion of the torus onto a rotationally supported supramassive neutron star remnant (i.e. a millisecond magnetar) powers two relativistic jetted outflows \citep{2012MNRAS.419.1537B,2018ApJ...857...95M} that, via internal dissipation mechanisms, produce the initial $\approx0.3 \,$s hard prompt gamma-ray emission. At this stage, the accretion disk releases winds that largely dominate the total mass ejected \citep{2019ApJ...880L..15M}. (c) The winds from the rotationally-powered magnetar are collimated by the surrounding ejecta, which give rise to the $\approx 45$-s duration soft gamma-ray emission \citep{2012MNRAS.419.1537B}. (d) As the spin-down luminosity of the magnetar decreases, the jetted winds become stifled behind the ejecta, which is reheated at larger radii. When the opacity of the ejecta decreases sufficiently, bright optical thermal emission is emitted \citep{2013ApJ...776L..40Y,2014MNRAS.439.3916M,2019LRR....23....1M}. (e) Hours after the merger, the ejecta is fully ionized by the winds of the long-lived magnetar, and the magnetar spin-down luminosity is detected \citep{2014MNRAS.439.3916M}.}
\label{fig:scheme}
\end{figure*}

%%%%%%%%%%%%%%%%%%%%%%%%%%%%%%%%%%%%%%%%%%%%%%%%%%%%%%%%%%%%%%%%%%%%%%%%%%%%%%%%

\section{Discussion}\label{sec:discuss}

The early-time multiwavelength observations of the short GRB 180618A propose a scenario in which only a long-lived magnetar remnant can account for all the observed emission components (see Figure~\ref{fig:scheme}): the extended soft gamma-ray emission following the short GRB \citep{2008MNRAS.385.1455M,2012MNRAS.419.1537B}, the unusual optical light curve \citep{2013ApJ...776L..40Y,2014MNRAS.439.3916M}, and the additional X-ray component \citep{2014MNRAS.439.3916M,2015ApJ...807..163G}.

Tens of magnetars have been identified in our Galaxy so far \citep{2017ARA&A..55..261K}, and some of them are regular X-ray bursters that, less frequently, emit at soft gamma-ray bands \citep{2021NatAs...5..372R}. More recently, giant flares from extragalactic magnetars have also been associated with low-luminosity short-duration GRBs \citep{2021Natur.589..211S}. However, the remnant of a neutron star binary merger is expected to be the more energetic version of a magnetar, a millisecond proto-magnetar \citep{2014MNRAS.439.3916M}, which is rotationally powered with typical energies $E_{\rm rot} \approx 10^{51}-10^{53} \,$erg, and will spin down until its collapse into a black hole \citep{2019ApJ...880L..15M}.

After a neutron star binary merger, if a newborn rapidly-spinning magnetar has sufficient spin-down luminosity, the winds will pierce through the ejecta and be collimated into bipolar jetted outflows that will dissipate Poynting-flux energy ---powering the extended gamma-ray emission of the GRB 180618A \citep{2008MNRAS.385.1455M,2012MNRAS.419.1537B}. As the spin-down luminosity of the magnetar decreases with $L_{\rm sd} \propto t^{-2}$, these winds are trapped behind the ejecta forming a hot nebula of electron-positron pairs that will radiate via synchrotron and inverse Compton emission \citep{2014MNRAS.439.3916M}. A fraction of the X-ray emission is then absorbed by the neutral ejecta walls, and reprocessed into optical and infrared photons that are able to escape when the optical depth of the expanding ejecta decreases enough. This allows the magnetar-powered kilonova of the GRB 180618A to be a hundred times brighter than a radioactively-powered kilonova \citep{2013ApJ...776L..40Y,2014MNRAS.439.3916M}. Hours to days after the burst, if the strong magnetar winds can completely ionize the ejecta, non-thermal X-ray emission will leak from the nebula producing an X-ray excess \citep{2014MNRAS.439.3916M}, similar to that observed at $\approx 0.5 \,$days after the GRB 180618A.

Our multiwavelength data also gives information about the geometry of the system. Given that we are detecting prompt gamma-ray emission that is bright and spectrally hard, we are likely facing the GRB jet \citep{2002ApJ...571L..31Y}. If the magnetar is releasing energy and accelerating ejecta along the polar regions of the system, material can easily reach trans-relativistic speeds \citep{2008MNRAS.385.1455M,2012MNRAS.419.1537B}. This extra kinetic energy is consistent with what we observe at optical bands; there is an early and rapid evolution of the thermal luminosity given the relativistically-expanding photospheric radius and the fast-fading spin-down luminosity of the magnetar, which we measure as $L_{\rm th} \propto t^{-(2.22\pm 0.14)}$. For the optical to be reprocessed within the observed timescales, we require an ejecta mass $M_{\rm ej } \lesssim 10^{-4} M_{\odot} (\kappa_{\pm}/10^3)^{-1} $ at the polar regions of the merger, which is reasonable given that a total ejected mass $\approx (0.01-0.3) M_{\odot}$ is expected in all directions \citep{2006MNRAS.368.1489O,2017ApJ...835L..34M}. This suggests that the merger ejecta distribution is considerably asymmetric, likely due to long-lasting cavities drilled by the early relativistic outflows or the disk winds ejecting more material in equatorial directions \citep{2012MNRAS.419.1537B}.

Current magnetohydrodynamic simulations cannot form jetted outflows just from the neutron star binary merger itself \citep{2017PhRvD..96h4063R}; successful jets require the formation and delayed collapse within a hundred milliseconds of an intermediate hypermassive neutron star \citep{2014ApJ...788L...8M}. However, constraints on the nuclear equation of state suggest that $18\%$ to $65\%$ of the neutron stars binary mergers will result in a less-massive and rotationally-supported supramassive neutron star remnant with longer lifetime \citep{2015ApJ...812...24F,2019ApJ...880L..15M}. Without the need for the neutron star remnant to collapse into a black hole, a viable short GRB from a merger could be powered by direct accretion onto the magnetar \citep{2012MNRAS.419.1537B}, or by the enhancement of the spin-down luminosity given the temporary presence of the accretion disk \citep{2018ApJ...857...95M}. Yet, baryon pollution remains a concern in these environments \citep{2007NJPh....9...17L,2014ApJ...788L...8M}.

The multiwavelength dataset of the GRB 180618A confirms GW170817/GRB 170817A findings \citep{2017ApJ...848L..13A} ---i.e. neutron star binaries as progenitors of short GRBs. While the remnant of the gravitational wave event GW170817 is likely a hypermassive neutron star that collapsed into a black hole within the first few hundreds of milliseconds after the merger \citep{2017ApJ...848L..13A,2019LRR....23....1M}, observations of the short GRB 180618A suggest a different outcome. We observe that a vast energy reservoir is injected into the system on timescales much larger than the duration of the accretion disk outflows ---powering several emission components across the spectrum that can only be explained by a long-lived magnetar remnant. Furthermore, it suggests that supramassive neutron stars with delayed collapse into a black hole are remnants of neutron star binary mergers \citep{2015ApJ...812...24F,2019ApJ...880L..15M}, and can power short-hard GRBs and extended soft gamma-ray emission through accretion and spin-down luminosity \citep{2008MNRAS.385.1455M,2012MNRAS.419.1537B}. These findings preserve a good agreement between the percentage of short GRBs that have extended emission ($13\%-50\%$; \citealt{2010ApJ...717..411N,2016ApJ...829....7L}), and the expected number of remnants from neutron star binary mergers that can power such emission ($18\%-65\%$; \citealt{2019ApJ...880L..15M}).

Future early-time studies of short GRBs with extended gamma-ray emission and joint GW/GRB detections will be able to statistically constrain how long and how many of these cosmological magnetars survive the merger, characterize the asymmetries in the distribution of the ejected mass, and probe jet acceleration in millisecond magnetars.

%%%%%%%%%%%%%%%%%%%%%%%%%%%%%%%%%%%%%%%%%%%%%%%%%%%%%%%%%%%%%%%%%%%%%%%%%%%%%%%%

\section{Conclusions}\label{sec:conclusion}

We report the multiwavelength observations of the short GRB 180618A; a GRB with unique gamma-ray, X-ray and optical properties result of a compact object binary merger at the outskirts of a galaxy at redshift $z=0.554 \pm 0.001$.

The bright prompt gamma-ray emission of the GRB 180618A consists of a multi-peaked structure with total duration $\approx 0.3\, $s and maximum energy radiated in the MeV domain, making the GRB 180618A one of the most energetic gamma-ray pulses ever detected among short-duration GRBs (i.e. flux, fluence, $E_{\rm peak}$). After the typically short and spectrally-hard gamma-ray pulse, we also detect a period of weak extended gamma-ray emission below $\approx 100 \,$keV, lasting $\approx 45 \,$s.

We find no detectable polarization at optical bands, and a rate of change of the light that initially follows a power-law $F_{\nu} \propto t^{-\alpha}$, with index $\alpha=0.46 \pm 0.02$. The optical emission is surprisingly short-lived, and the slow decline is replaced by a sudden drop in brightness at $35 \, $minutes post-burst, steepening to $\alpha=4.6 \pm 0.3$. The light curve break progressively passes from the ultraviolet to near-infrared bands. Afterwards, there is no further detection of the optical transient at the GRB 180618A coordinates.

The GRB 180618A optical counterpart presents temporal and spectral properties that do not satisfy the characteristic scalings of the synchrotron spectrum of the GRB afterglow \citep{1998ApJ...497L..17S} ---powered by the shock of the relativistic collimated ejecta with the circumburst medium. In contrast, the fast-fading X-ray emission is consistent with a decelerating jetted outflow, and with an extra emission component at $\approx 0.5 \,$days post-burst. This leads us to consider two distinct mechanisms powering the X-ray and unusual optical emission.

The modelling of the overall emission suggests thermal-like emission from a relativistically-expanding source dominating the optical emission at $\approx 15-60 \,$minutes post-burst, which naturally accounts for the sharp chromatic drop of the optical emission at high-frequency wavelengths. Furthermore, the X-ray to optical emission before $15 \,$minutes post-burst is consistent with the fast-fading jet afterglow.

We interpret the unusual spectral and temporal properties of the GRB 180618A as evidence of a highly magnetized, spinning neutron star that survives for longer than $\approx 10^5 \,$s after the merger, and spins down at rate $L_{\rm th} \propto t^{-(2.22\pm 0.14)}$ powering a relativistically-expanding hot thermal nebula in the process. Here, we confirm that newborn millisecond magnetars can power bright emission components across the electromagnetic spectrum that remain detectable at cosmological distances: i.e. the extended soft gamma-ray emission following some short GRBs \citep{2008MNRAS.385.1455M,2012MNRAS.419.1537B}, optical plateaus at early times \citep{2017A&A...607A..84K}, the fast-evolving bright thermal optical emission \citep{2013ApJ...776L..40Y,2014MNRAS.439.3916M} and the late-time flattening of the X-ray light curve \citep{2014MNRAS.439.3916M,2015ApJ...807..163G}. The early afterglow emission drop, and the short-lived thermal optical emission may explain why such thermal emission has not been detected yet in other short GRBs with extended emission; this discovery opens a new era for searches of gravitational wave counterparts with fast-cadence surveys.

%%%%%%%%%%%%%%%%%%%%%%%%%%%%%%%%%%%%%%%%%%%%%%%%%%%%%%%%%%%%%%%%%%%%%%%%%%%%%%%%

\begin{acknowledgments}

We thank the anonymous referee for their constructive comments that improved the clarity and accuracy of the paper. The research leading to these results has received funding from the European Union's Horizon 2020 Programme under the AHEAD project (grant agreement 654215). N.J.-M. and C.G.M. acknowledge financial support from Mr Jim Sherwin and Mrs Hiroko Sherwin. C.M. acknowledges support from the Science and Technology Facilities Council and the UK Research and Innovation (ST/N001265/1). E.R-R.  is supported in part by NASA grant NNG17PX03C, NSF grant AST-1911206, AST-1852393, and AST-1615881, and the Heising-Simons Foundation. A.G. acknowledges the financial support from the Slovenian Research Agency (grants P1-0031, I0-0033, J1-8136, J1-2460). M.M. acknowledges financial support from the Italian Ministry of University and Research - Project Proposal CIR01\_00010. We acknowledge A. Becker for taking the data, R.T. Gatto for useful discussions, and D. Paris for the help in the data reduction of the LBC. We thank E. Burns and D. Burrows for useful discussions.

\end{acknowledgments}

\vspace{5mm}
\facilities{Swift (BAT, XRT and UVOT), Fermi (GBM), Liverpool:2m (RINGO3 and IO:O), LBT (LBC and MODS)}

\software{Matplotlib \citep{2007CSE.....9...90H}, SciPy \citep{2020NatMe..17..261V}, PyFITS \citep{1999ASPC..172..483B}, Astropy \citep{2013A&A...558A..33A,2018AJ....156..123A}, Astropy Photutils \citep{2016ascl.soft09011B}, Astroalign \citep{2019arXiv190902946B}, Xspec and PyXspec (v12.9.1; \citealt{1996ASPC..101...17A,1999ascl.soft10005A}), HEAsoft (v6.22.1; \citealt{1995ASPC...77..367B}), RMFit (v4.3.2; \citealt{2014ascl.soft09011G})}

\bibliography{biblio}{}
\bibliographystyle{aasjournal}

\begin{deluxetable}{cccc}[ht!]
\tablecaption{The GRB 180618A optical photon indexes ($\beta_{\rm opt, PI}$) derived from the best-fitting power-law models to the RINGO3 data. \label{tab:PL}}
\tablecolumns{8}
\tablewidth{0pt}
\tablehead{
\colhead{t$_{\rm mid}$-T$_0$} & \colhead{t$_{\rm err}$}  & \colhead{$\beta_{\rm opt, PI}$} & \colhead{$\beta_{\rm opt, PI, \, err}$} \\
 \colhead{(s)} & \colhead{(s)} & \colhead{} &  \colhead{}
}
\startdata
	231 &       30 &      0.7 &      0.4 \\
     340 &       42 &      1.2 &      0.4 \\
     473 &       89 &      1.2 &      0.4 \\
     749 &      179 &      1.3 &      0.4 \\
     995 &       67 &      1.5 &      0.4 \\
    1212 &      150 &      1.4 &      0.4 \\
    1528 &       90 &      1.2 &      0.4 \\
    1825 &      208 &      1.8 &      0.5 \\
    2670 &      298 &      2.6 &      0.6 \\
    3288 &      298 &      4.0 &      0.8 \\
\enddata
\tablecomments{The t$_{\rm mid}$ is the mean observing time, the T$_{0}$ is the BAT trigger time, and the t$_{\rm err}$ is half the length of the observing time window. Note that the model does not account for host galaxy extinction.}
\end{deluxetable}

\begin{deluxetable}{cccccc}[ht!]
\tablecaption{The GRB 180618A effective temperatures ($T_{\rm eff}$) and luminosities ($L_{\rm th}/D_{\rm L}^2$) derived from the best-fitting black body models to the joint RINGO3/UVOT data. \label{tab:BB}}
\tablecolumns{8}
\tablewidth{0pt}
\tablehead{
		\colhead{t$_{\rm mid}$-T$_0$} & \colhead{t$_{\rm err}$}  & \colhead{$T_{\rm eff}$} & \colhead{$T_{\rm eff, \, err}$}  & \colhead{$L_{\rm th}/D_{\rm L}^2$} & \colhead{$L_{\rm th, err}/D_{\rm L}^2$} \\
		 \colhead{(s)} & \colhead{(s)} & \colhead{($10^3 \,$K)} & \colhead{($10^3 \,$K)} & \colhead{(10$^{43}$ erg s$^{-1}$ Gpc$^{-2}$)} & \colhead{(10$^{43}$ erg s$^{-1}$ Gpc$^{-2}$)} 
}
\startdata
231 &       30 &     10.9 &      1.0 &     42.6 &      5.3\\
     340 &       42 &      9.8 &      0.7 &     27.9 &      2.2\\
     473 &       89 &     10.0 &      0.5 &     27.9 &      0.7\\
     749 &      179 &      9.8 &      0.6 &     20.7 &      1.7\\
     995 &       67 &     10.2 &      0.6 &     19.8 &      1.0\\
    1212 &      150 &      8.6 &      0.9 &     13.0 &      1.5\\
    1528 &       90 &      7.8 &      1.1 &      9.4 &      1.2\\
    1825 &      208 &      7.2 &      1.2 &      7.1 &      0.9\\
    2670 &      298 &      5.3 &      0.9 &      2.6 &      0.3\\
    3288 &      298 &      3.8 &      0.6 &      1.8 &      0.6\\
\enddata
\tablecomments{The t$_{\rm mid}$ is the mean observing time, the T$_{0}$ is the BAT trigger time, the t$_{\rm err}$ is half the length of the observing time window, and $D_{\rm L}$ is the luminosity distance. Note that the model does not account for host galaxy extinction.}
\end{deluxetable}

\end{document}